\documentclass[iop]{emulateapj}

\usepackage{graphicx}
\usepackage{natbib}
\usepackage{amsmath}
\usepackage{amssymb}
\usepackage{fancyhdr}
\usepackage{latexsym}
\usepackage{color}
\usepackage{txfonts}
\usepackage{subfigure}

  \received{\today}
  \revised{}
  \accepted{}
  
\begin{document}

   \title{COMBINED MODELING OF ACCELERATION, TRANSPORT, AND HYDRODYNAMIC RESPONSE IN SOLAR FLARES. II. Inclusion of Radiative Transfer with RADYN}

   \author{Fatima Rubio da Costa\altaffilmark{1}, Wei Liu\altaffilmark{1}$^,\,$\altaffilmark{2}, Vah\'{e} Petrosian\altaffilmark{1} and Mats Carlsson\altaffilmark{3}}

\altaffiltext{1}{Department of Physics, Stanford University, Stanford, CA 94305, USA; Email: frubio@stanford.edu}
\altaffiltext{2}{W.~W.~Hansen Experimental Physics Laboratory, Stanford University, Stanford, CA 94305, USA}
\altaffiltext{3}{Institute of Theoretical Astrophysics, University of Oslo, P.O. Box 1029 Blindern, N-0315 Oslo, Norway}

\begin{abstract}
Solar flares involve complex processes that are coupled and span a wide range of temporal, spatial, and energy scales. Modeling such processes self-consistently has been a challenge in the past.
Here we present results from simulations that couple particle kinetics with hydrodynamics of the atmospheric plasma. We combine the Stanford unified Fokker-Planck code that models particle acceleration and transport with the RADYN hydrodynamic code that models the atmospheric response to collisional heating by accelerated electrons through detailed radiative transfer calculations. We perform simulations using two different electron spectra, one an {\it ad hoc} power law and the other predicted by the model of stochastic acceleration by turbulence or plasma waves. Surprisingly, the later model, even with energy flux $\ll 10^{10}$~erg~s$^{-1}$~cm$^{-2}$, can cause ``explosive" chromospheric evaporation and drive stronger up- and downflows (and hydrodynamic shocks). This is partly because our acceleration model, like many others, produces a spectrum consisting of a quasi-thermal component plus a power-law tail.
We synthesize emission line profiles covering different heights in the lower atmosphere, including H$\alpha$~6563~\AA, \ion{He}{2}~304~\AA, \ion{Ca}{2}~K~3934~\AA\ and \ion{Si}{4}~1393~\AA. One interesting result is the unusual high temperature (up to a few $10^5$~K) of the formation site of \ion{He}{2}~304~\AA, which is expected due to photoionization-recombination under flare conditions, compared to those in the quiet Sun dominated by collisional excitation.
When compared with observations, our results can constrain the properties of non-thermal electrons and thus the poorly understood particle acceleration mechanism.
\end{abstract}

\keywords{acceleration of particles --- hydrodynamics (HD) --- line: profiles --- radiative transfer --- Sun: flares; chromosphere}

   \section{Introduction}\label{intro}
One of the outstanding problems in solar physics is how magnetic energy is transformed into the observed signatures of solar flares. It has been recently recognized that modeling the coupling between the particle acceleration and transport processes in solar flares and the dynamical response of the atmosphere to particle collisional heating is critical to our understanding of flare dynamics. It is clear that non-thermal electrons and ions play an important role. However, the exact mechanism of acceleration of these particles is still a matter of considerable debate. Several scenarios have been proposed and different models have been developed with different degrees of details. Among these are acceleration by DC electric fields \citep[e.g.,][]{1985ApJ...293..584H}, shocks \citep[e.g.,][]{1998ApJ...495L..67T}, and turbulence.
Stochastic acceleration by turbulence or plasma waves \citep{1979AIPC...56..135R, 1992ApJ...398..350H, 1997ApJ...491..939M} has been developed in greater detail \citep{2004ApJ...610..550P} and has been tested by observations more rigorously \citep{1999ApJ...527..945P, 2004ApJ...613L..81L, 2006ApJ...636..462L}; for a recent review see \cite{2012SSRv..173..535P}.

Observations most intimately connected to the acceleration process are the  microwave, Hard X-ray (HXR) and $\gamma$-ray radiations. These emissions are produced by particles, which are (most likely) accelerated in the corona in a flaring loop. The accelerated electrons produce microwave and HXR emissions via synchrotron and bremsstrahlung processes \citep{1981ApJ...246L.155H, 1994PhDT.......335S, 1994ApJS...90..599K}, while the interaction of accelerated protons and ions with the background ions produce $\gamma$-rays \citep{1985SoPh..100..537L}. However, most of the energy of the particles goes into heating of the plasma via Coulomb collisions as they travel down to the footpoints. The heating and evaporation of the plasma produces secondary continuum and line emissions, from infrared to soft X-ray energy bands, which also carry information about the acceleration process. The purpose of the work presented here is to explore this second avenue of testing the acceleration models. 

For the long run, our aim is to use the combined non-thermal and thermal signatures to distinguish between the different acceleration scenarios and constrain the characteristics of the specific acceleration models. This requires a combined treatment of the acceleration, transport and radiation of particles and hydrodynamic (HD) response of the atmosphere to the energy input by particles. The first numerical study of particle transport in solar flares was carried by \citet{1981ApJ...251..781L} who treated the transport and radiation by electrons using the Fokker-Planck transport equation taking into account the pitch angle changes due to Coulomb collisions and magnetic field variations. This study was extended by \citet{1990ApJ...359..524M}, who considered energy loss and pitch angle changes due to synchrotron emission. This transport-radiation code was later combined with an stochastic acceleration code \citep{1995ApJ...446..699P, 1999ApJ...527..945P, 2004ApJ...610..550P} into the unified Stanford code \citep{2002ApJ...569..459P}.

The HD response has been investigated in several works \citep{1989ApJ...341.1067M, 2009A&A...499..923K}  by the means of 1D numerical hydrodynamics along a coronal loop. These works assumed that non-thermal electrons with a power-law spectrum were injected at the apex of the loop and used to approximate analytic expressions for the transport and energy deposition along the loop. \citet{1999ApJ...521..906A} and \citet{2005ApJ...630..573A} improved the results by including a detailed calculation of radiative transfer in the atmosphere.

We have embarked on development of more complete and self-consistent treatment of this important problem. Instead of using an ad-hoc power-law injected electrons and approximate treatment of transport, we have combined the Stanford Fokker-Planck acceleration-transport code with HD codes, achieving a more accurate determination of the radiative signatures of flares. In our first paper, \citet{2009ApJ...702.1553L} (hereafter {\bf paper I}), we combined the Stanford code with the Naval Research Laboratory (NRL) hydro-code \citep{1989ApJ...341.1067M}, where we addressed some of the non-thermal aspects of the problem. Here we extend this study by taking into account the detailed calculation of radiative transfer in the atmosphere. Instead of the NRL code we use the radiative HD (RADYN) code \citep{1992ApJ...397L..59C, 1997ApJ...481..500C}, in a modified version \citep{1999ApJ...521..906A, 2005ApJ...630..573A}, and focus on intensities and shapes of several lines emitted at different heights. The main goal here is to show the effects of more realistic accelerated spectrum of electrons as compared to those of power-law injected electrons. In a recent paper we have used the RADYN code to interpret spectroscopic observations of IR-optical-UV continuum and line emissions \citep{2015ApJ...804...56R}. More such comparison with specific observations, both line and (thermal and non-thermal) continuum emissions  will be presented in future publications using this combined code. 

This paper is organized as follows. In Section~\ref{sect:method} we describe our model and the method used to solve the combined radiative HD and Fokker-Planck equations. In Section~\ref{sect:compare_runs1_2}, we present results on plasma characteristics from the simulations and compare them with those from the power-law injection case. In Section~\ref{sect:lines_emission} we focus on the effects of electron heating on the emission in the H$\alpha$, \ion{Ca}{2}~K, \ion{He}{2}~304~\AA\ and \ion{Si}{4}~1393~\AA\ lines. A brief summary and conclusions are given in Section~\ref{Sect:conclusion}. In Appendix~\ref{sect:internal_energy} we present details of different energy terms and their influence on the atmospheric evolution and in Appendix~\ref{sect:results_flux} we investigate the effects of different electron fluxes.

\section{Model and Method}\label{sect:method}   
  \subsection{Assumptions and Geometry}
Magnetized plasma processes, in general, are commonly modeled with multidimensional magnetohydrodynamics (MHD) codes. However, in the solar corona the magnetic pressure dominates over the gas pressure (i.e. the plasma  $\beta\ll 1$), so plasma flows mainly along magnetic field lines. This allows us to use the HD instead of MHD treatment of the problem. 

We assume a one-dimensional, semi-circular loop perpendicular to the solar surface with a constant diameter of 3~Mm. Adopting a symmetric boundary condition at the loop apex, we construct a computational domain containing a 10.4~Mm long quarter circle, discretized in 191 grid points (a denser grid would be computationally too expensive). The loop extends in a plane-parallel model atmosphere from the corona to the bottom of the chromosphere at $z=0$~Mm, where the optical depth $\tau_{5000}$ at a wavelength of $\lambda=5000 \,\, \AA$ is unity. The circular geometry is taken into account when calculating the X-ray photoionization rates and the gravitational acceleration; otherwise the loop is treated as a vertical cylinder.

For the pre-flare conditions we use the FP2 model of \citet{1999ApJ...521..906A}, which was generated by adding a transition region and corona to the model atmosphere of \citet{1997ApJ...481..500C}. The temperature is fixed to 10$^6$ K at the loop apex. By running the code without external heating, the atmosphere relaxes to a state of HD equilibrium.

   \subsection{Radiative Transfer and Hydrodynamics}
The \begin{bf}RADYN\end{bf} code simultaneously solves the equations of hydrodynamics, population conservation, and radiative transfer using a one-dimensional adaptive grid \citep{1987JCoPh..69..175D}.

Atoms important to the chromospheric energy balance are treated in non-local thermodynamical equilibrium (non-LTE). These include a six-level plus continuum hydrogen atom; a six-level plus continuum, singly ionized calcium atom; a nine-level plus continuum helium atom; and a four-level plus continuum, singly ionized magnesium atom. The transitions that are treated in detail are given in Table 1 of \citet{1999ApJ...521..906A}. Complete redistribution is assumed for all lines, except for the Lyman series in which partial frequency re-distribution is mimicked by truncating the profiles at 10 Doppler widths \citep{1973ApJ...185..709M}. Other atomic species are included in the calculation as background continua in LTE, using the Uppsala opacity package of \citet{Gustafsson}.

\begin{figure}[!hbt]
 \centering
 \epsscale{1.1}
 \plotone{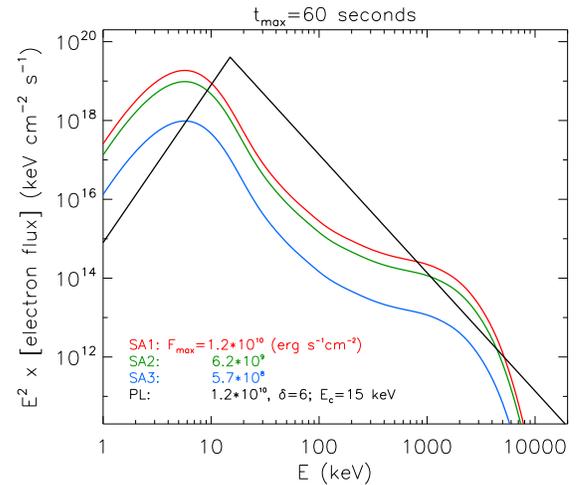}
\caption{Spectra of angle-integrated electron fluxes, $F(E)$, times $E^2$ at the loop-top and $t_{max}$=60~s for different simulation runs labeled with the corresponding energy fluxes $\mathfrak{F}_{max}$. The black line represents a power law with a spectral index $\delta=5$ and low energy cutoff $E_c=15$ keV (Run~PL), while the colored lines represent stochastically accelerated electron spectra (Runs~SA1--SA3). [A color version of this figure is available in the online journal.].}
 \label{e_spectra_top}
\end{figure}

 \begin{table*}[!htb]
 {\centering
 \caption{Summary of the different simulation runs and the atmospheric parameters for different flaring conditions.}
 \label{table_runs}
 \tabcolsep 0.15in
 \begin{tabular}{c c c c c c c c c}
   \hline \hline
  Run & Injected e$^-$ distribution & $\mathfrak{F}_{\rm max}$   & $v_{\rm max}$ & $t$($v_{\rm max}$) & $v_{\rm min}$ & $t$($v_{\rm min}$)& $T_{\rm max}$ & $t$($T_{\rm max}$)\\  & & (erg~s$^{-1}$~cm$^{-2}$) & (km s$^{-1}$) & (s) & (km s$^{-1}$) & (s) & (K) & (s)\\  \hline
  PL & $\delta$=5; $E_c$=15 keV & 1.2$\times10^{10}$ & 483 & 46 & $-44$ & 30 & $9.5 \times 10^6$ & 75\\
  SA1 & Stochastic acceleration & 1.2$\times10^{10}$ & 750 & 32 & $-41$ & 10 & $2.3 \times 10^7$ & 60\\
  SA2 & Stochastic acceleration & 6.2$\times10^9$ & 641 & 37 & $-39$ & 13 & $1.8 \times 10^7$ & 60\\
  SA3 & Stochastic acceleration & 5.7$\times10^8$ & 377 & 77 & $-28$ & 31 & $8.0 \times 10^6$ & 61\\
  \hline
 \end{tabular}
\\
{Notes: $\mathfrak{F}_{\rm max}$ -- maximum electron energy flux. $v_{\rm max}$ \& $t (v_{\rm max})$, $v_{\rm min}$ \& $t (v_{\rm min})$, and $T_{\rm max}$ \&  $t (T_{\rm max})$ are the maximum velocity (upflow, $v>0$), minimum velocity (downflow, $v<0$), and maximum temperature and their corresponding times, respectively.}
 }
\end{table*}

 \begin{figure*}[!htb]
 \centering
 \epsscale{1.15}
  \plotone{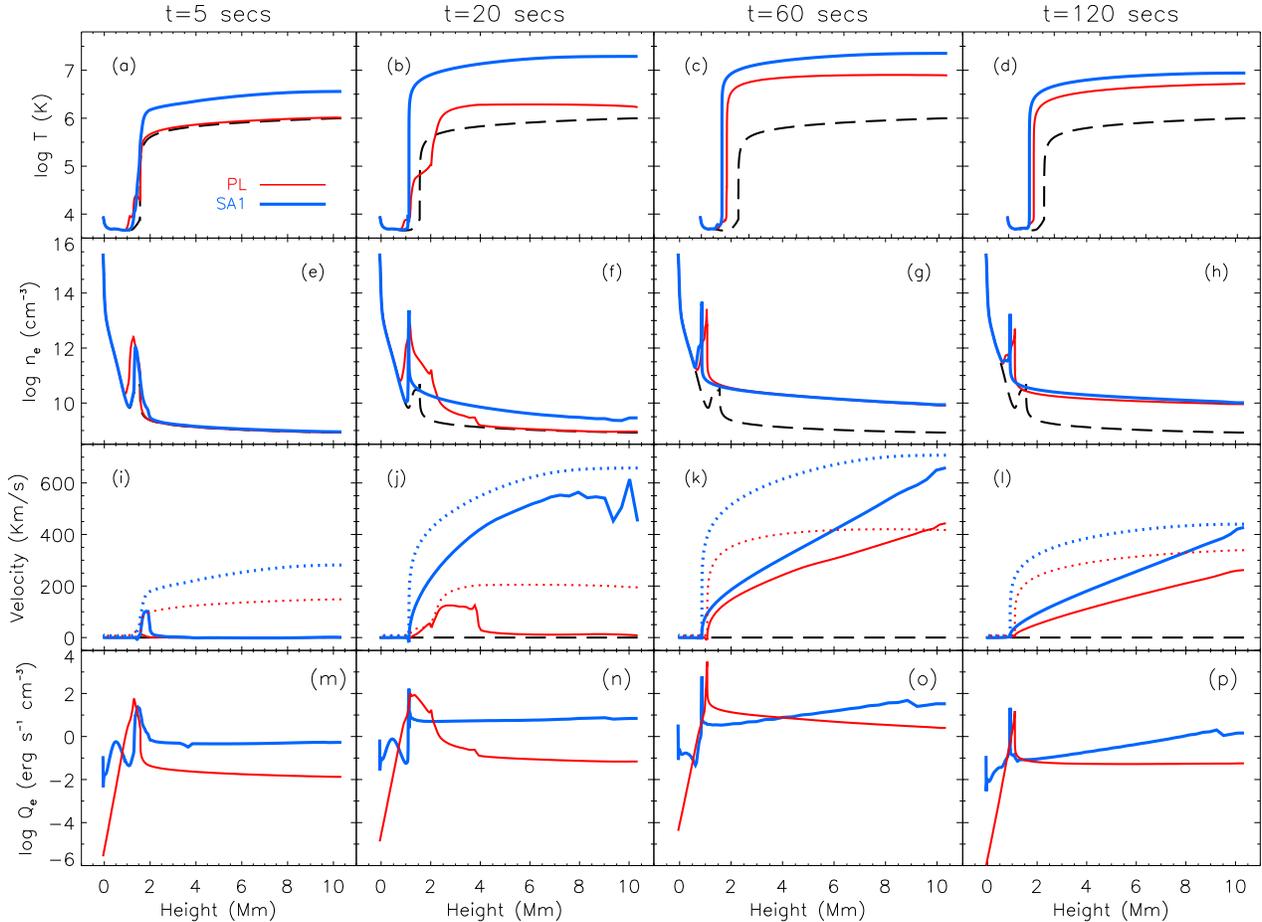}
  \caption{Variation with distance along the loop of the atmospheric parameters (temperature, electron density, velocity of the plasma and electron heating rate per unit volume ($Q_e$) where positive velocity refer to upflows and negative velocities to downflows) in the PL (red line) and SA1 (blue line thick) models. The black dashed line show the initial values and the dotted lines in the bottom panels represent the sound speed. The columns are for $t=5$, 20, 60 and 120~s. (see the online movie for more details.). [A color version of this figure is available in the online journal.].} 
  \label{evol_atm}
\end{figure*}

As described in \citet{2005ApJ...630..573A}, the radiative HD code includes the calculation of photoionization heating resulting from high temperature, soft X-ray emitting regions, as well as the calculation of optically-thin cooling due to thermal bremsstrahlung and collisionally excited metal transitions. An adjustment in the calculation of the conductive flux has also been taken into account in order to avoid unphysical large values in the transition region -- where temperature gradients are large. Hydrodynamic effects due to gravity, thermal conduction, and compressional viscosity are considered as described in \citet{1999ApJ...521..906A}.

Inclusion of the radiative transfer calculation gives us the advantage, with respect our Paper~I, to investigate how the electron deposition in the chromosphere affects to the emission lines originating from  several heights (see Section~\ref{sect:lines_emission}).
 

   \subsection{Electron Acceleration, Transport and Energy Deposition}\label{SAmodel}
{\bf FLARE} -- the Stanford unified acceleration-transport code \citep{1990ApJ...359..524M, 2004ApJ...610..550P} consists of two modules. The acceleration module calculates the spectrum of the electrons accelerated stochastically by turbulence in the acceleration region \citep[assumed to be located at the loop-top; for observational evidence see, e.g.,][]{2013ApJ...767..168L} and the spectrum of electrons escaping down to the footpoints. The transport module then calculates the evolution of the spectrum and pitch angle distribution along the loop of the escaping electrons from the acceleration site. The code includes energy and pitch-angle diffusion due to Coulomb collisions and synchrotron radiation. 

One of the main inputs to the RADYN code is the heating rate $Q_e(s)$ (in units of erg~s$^{-1}$~cm$^{-3}$) due to accelerated electrons as a function of position $s$ along the loop, which is included as a source of external heating in the energy conservation equation (see Equation~\ref{eq_energy_conservation} in Appendix~\ref{sect:internal_energy}). \citet{1999ApJ...521..906A} assumed a beam of electrons with a power-law spectrum with an index $\delta$ and low energy cutoff $E_c$ and used the approximate analytic expression of \citet{1981ApJ...249..817E} to calculate the heating rate. Here, we use the electron spectrum from the acceleration-transport code, which in general includes a quasi-thermal plus a non-thermal component. We then calculate the heating rate $Q_e(s)$ using the spatial variation of the spectrum and pitch angle distribution $f(E,\mu, s)$ of the electrons, following the procedure described in Section~3.1 of Paper~I.

   \subsection{Combining FLARE and RADYN}
We followed the approach detailed in Paper~I to combine the FLARE and RADYN codes. In brief, we note that the current particle transport module of FLARE can provide only a steady-state solution. To perform time-dependent simulation, RADYN calls the transport module every $\Delta t=1$~s. This interval of communication between the two codes allows for a steady-state solution of particle kinetics, whose timescales are orders of magnitude shorter than HD timescales. This interval also provides a trade-off for computational efficiency.

In each iteration, the RADYN code passes the electron density of the atmosphere as a function of distance from the top of the loop ($n_e(s)$) to the FLARE code. The FLARE code calculates the electron heating rate $Q_e(N)$ as a function of column density $N_e= \int_0^s n_e(s) \; ds $ and return it to the RADYN code, which then simulates the evolution of the atmosphere and advance in time for $\Delta t=1$~s. Within this interval, we assume that the accelerated electron flux remains constant and so does $Q_e(N_e)$. Therefore, the temporal evolution of the heating rate $Q_e(s, \, t)$ as a function of distance solely depends on the spatial redistribution of the electron density $n_e(s, \, t)$ with time, as described in Section~3.2 of Paper~I.

By including the radiative transfer calculation, one of the advantages with respect our previous Paper~I is the study in detail how the electron deposition in the chromosphere affects to the emission of the lines at several heights (see Section~\ref{sect:lines_emission}).

   \subsection{Simulation Runs}\label{sect:runs}
We have performed four simulation runs with different injected electron spectra, as summarized in Table \ref{table_runs}, to investigate their different atmospheric response. These runs were carefully designed and allowed us to compare our results using the acceleration model based spectra with those carried out in the past using a simple power-law. In the Run~PL, we use a single power-law electron spectrum of index $\delta$=5 and low energy cutoff of $E_c$=15 keV and the analytic expression of \citet{1978ApJ...224..241E, 1981ApJ...249..817E} to calculate the electron heating rate $Q_e$. Runs~SA1--SA3 use the result of the stochastic acceleration (SA) model for the injected electron spectra, for which we prescribed a common characteristic acceleration timescale of $\tau_p$ = 1/70~s (\citealt{2004ApJ...610..550P}) and thus the same spectral shape. For the SA runs we calculate the $Q_e$ using the approach described in Section~\ref{SAmodel}. 
Note that Run~SA1 has identical electron spectra with Run~N in Paper~I and Run~PL has a lower spectral index than Run~O in Paper~I. Each simulation lasts 120~s with the electron energy flux $\mathfrak{F}(t)$ increasing linearly till $t_{max}$=60~s up to $\mathfrak{F}_{max}$ and then decreasing for another 60~s. In all the runs we keep the electron spectral shape constant in time. 

Figure~\ref{e_spectra_top} compares the spectra of angle-integrated electron flux at the top of the loop for the different runs shown in Table~\ref{table_runs}. In the spectral range approximated for hard X-ray bremsstrahlung emission, the slope of the PL spectrum has been chosen to be similar to the one of the SA~Run. The main difference between the two models is  the absence of the quasi-thermal electrons in the PL model and the secondary difference is the roll over in the SA model at highest energies.

We should, however, note that these differences are not unique to the SA model. Most models, when accelerating particles from a thermal pool, produce a hotter quasi-thermal component with a power law tail because of Coulomb collision effects. Interested readers are referred to, for example, \citet{2008ApJ...682..175P} for a generic acceleration model and \citet{2009ApJ...707L..92S} for shock acceleration using particle-in-cell (PIC) simulations. We also note that the selection of the spectral parameters has not been done to study any particular flare.

In the next section we investigate how the atmosphere responds to the injection of single power law electrons (Run~PL) and stochastically accelerated electrons (Run~SA1) and in Appendix~\ref{sect:results_flux} we compare results of Runs~SA1, SA2 and SA3 to investigate how the atmosphere responds to the variation of the electron flux. 

   \section{Comparison of single power-law injection with stochastic acceleration}\label{sect:compare_runs1_2}

\begin{figure}[!hbt]
\centering
  \epsscale{1.2}
  \plotone{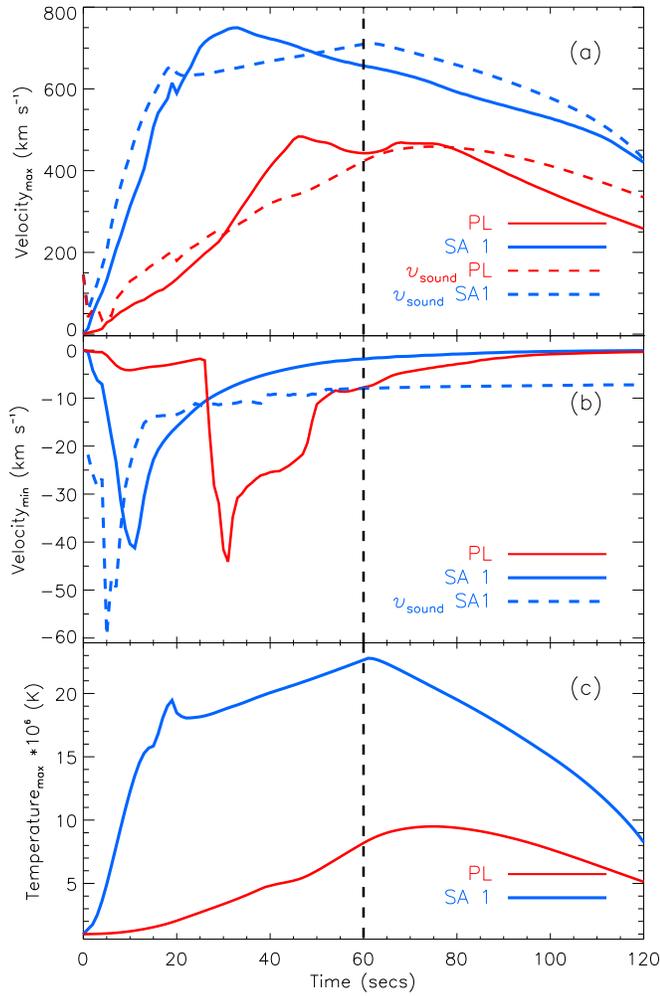}
  \caption{Temporal evolution of the maximum (a) upflow velocity, (b) downflow velocity, and (c) temperature within the loop for Runs~PL (red line) and SA1 (blue line thick). The dashed curve indicates the sound speed at the corresponding location of the maximum upflow (a) and downflow (b) velocity for each run. [A color version of this figure is available in the online journal.].}
     \label{vel_max}
\end{figure}

\begin{figure*}[!bth]
\centering
  \epsscale{1.05}
  \plotone{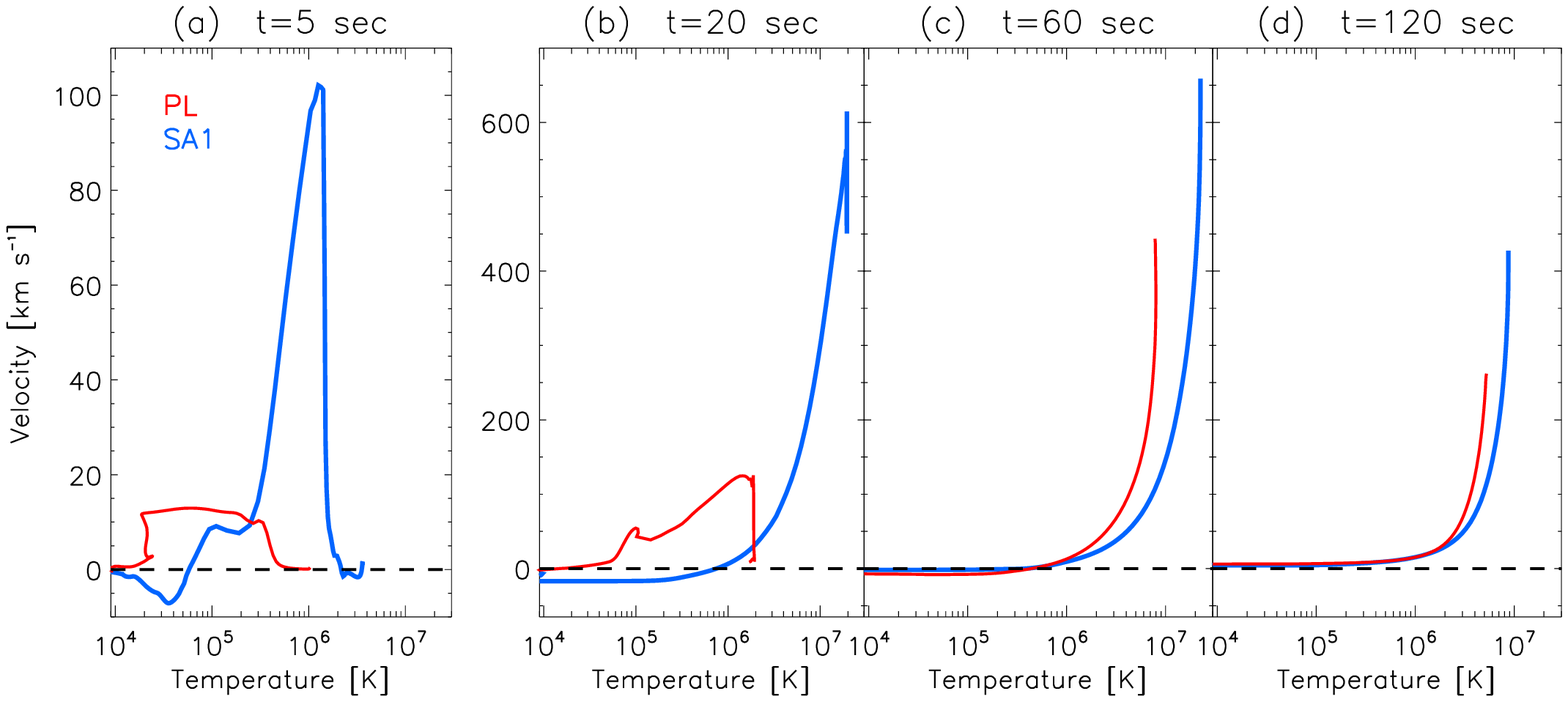}
  \caption{Temporal evolution of the velocity vs. temperature for PL (red solid line) and SA1 (blue solid thick line). [A color version of this figure is available in the online journal.].}
     \label{vel_temp}
\end{figure*}

In order to investigate how the acceleration and transport of electrons affects the atmospheric response, we compare the results of injection of single power-law electrons (Run~PL) and of stochastically accelerated electrons (Run~SA1), which have the same total energy flux at any moment.

    \subsection{Atmospheric evolution}
Figure~\ref{evol_atm} and the online movie shows the temporal evolution of the atmosphere. In general SA1 has higher values than PL for all the atmospheric variables (see Table~\ref{table_runs}), qualitatively consistent with those of Runs~N and O in Paper~I, respectively. For example, the corona is heated more rapidly to higher temperatures in SA1 than that in PL. The maximum coronal temperature of $2.3 \times 10^7$~K is reached at $t=60$~s in SA, compared with a lower value of $9.5 \times 10^6$~K at 74~s in PL (Figure~\ref{vel_max}(c)). As detailed in Section~\ref{e-heating}, such contrasts are due to the different spatial distributions of collisional heating by non-thermal electrons of the two models, even though they have the same energy input to the loop as a whole. This factor of 2.4 difference in the maximum coronal temperature is much greater than the factor of 1.2 difference between similar runs in Paper~I (see their Table~1), which points to the importance of detailed radiative transfer calculation included here.

At early times, electron heating in the transition region and chromosphere causes an overpressure that drives both upflows (i.e., chromospheric evaporation) and downflows of plasma, with higher speeds in the case of SA1. At $t=5$~s, for example, SA1 has the maximum upflow and downflow velocities of 102 and $-7$~km~s$^{-1}$, respectively, compared with 13 and $-0.4$~km~s$^{-1}$ in PL (see Figure \ref{vel_max} for the temporal evolution). As a result of the upflow, part of the initial transition-region and chromospheric material is converted into coronal mass, and the new transition region recedes to lower heights between $z=0.99$ and 1.10~Mm.

At $t=5$~s most of the energy is deposited in a narrow region located in the upper chromosphere at a height of $z=1.27$~Mm for PL and at 1.34~Mm for SA1. The non-thermal electrons quickly heat this region to temperatures greater than 10$^4$ K. The location of the maximum non-thermal energy deposition gradually moves downwards, reaching a height of 1.17~Mm for PL and $1.12$~Mm for SA1 at $t=20$~s. A more detailed discussion on how the electron heating rate affects the atmosphere is given in Appendix~\ref{sect:internal_energy}, where the height distribution of the different energy terms and their temporal evolution are presented. We note that in the case of PL a significant fraction of hydrogen is ionized at $t=5$~s and $z=1.18$~Mm and thus the energy can no longer be effectively radiated away by bound-bound transitions.

Figure~\ref{vel_max}(a) shows the temporal evolution of the maximum velocity along the loop (solid lines) and the sound velocity at the same position (dashed lines). Again, SA1 has higher plasma velocities, creating a hydrodynamic shock with supersonic speeds between $t=22$ and 49~s, while in the PL case, a shock is developed at later times between $t=30$ and 78~s. 
For SA1, the maximum upflow velocity of $v_{max}=750$~km s$^{-1}$ with a Mach number of 1.15 is attained at $t=32$~s, while for PL the maximum velocity $v_{max}=483$~km s$^{-1}$ with a Mach number of 1.41 occurs at $t=46$~s.

As shown in Figure~\ref{vel_max}(b), between $t=9$ and 24~s for the SA1~Run, the downflow plasma velocity exceeds the sound speed and forms a downward propagating shock. In contrast, the sound speed (not shown) for the PL case in (b) is alway greater than the maximum downflow speed, except for a brief period of 3~s from $t=32$ to $t=34$~s.

In Figure~\ref{vel_max}(c) we can also see that the temporal evolution of the maximum temperature along the loop follows a similar general trend as the electron flux $\mathfrak{F}(t)$, but the temperature in the SA1 run increases much faster than that in the PL run.

Figure~\ref{vel_temp} shows the temperature distribution of the plasma velocity at different times. The two runs exhibit distinct behaviors at early times. At $t=5$ and 20~s (panels (a) and (b)), for example, in the PL case, mainly low-speed upflows in the $10^4$--$10^6$~K range are present. In the SA1 case, downflows occur at low temperatures and high-speed upflows at high temperatures, with their division temperature increasing with time from a few $10^4$~K to nearly $10^6$~K. Comparably slower downflows take place in the PL case only at later times (e.g., 60~s). As time progresses toward the late phase of the flare, such a distinction diminishes to lesser degrees, with mainly upflows present in both cases whose speed rapidly grow with temperature.

\begin{figure*}[tbh]
\centering
  \epsscale{1.0}	
  \plotone{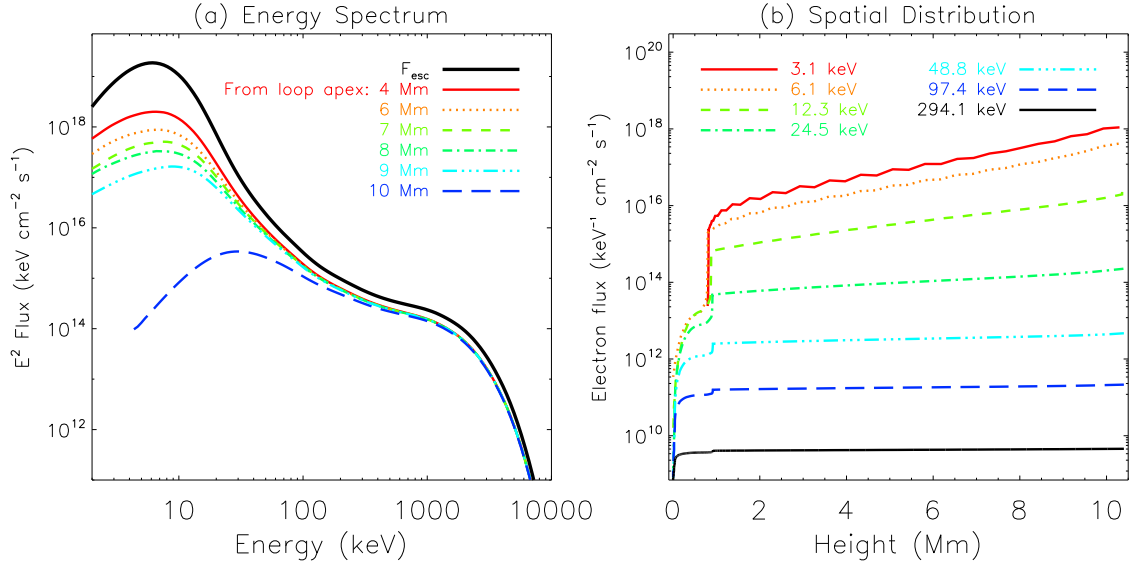}
  \caption{Energy and spatial distributions of non-thermal electrons at $t_{max}=60$~s in Run~SA1. (a) Energy spectra of angle-integrated electron flux multiplied by the energy squared. The solid black line is the flux ($F_{esc}$) of electrons escaping the acceleration region at the loop apex that serves as the injection to the transport portion of the loop. The color-coded curves (from red to blue) of different line styles represent fluxes at growing distances of 4, 6, 7, 8, 9, and 10~Mm from the loop apex.
  (b) Spatial distribution of angle-integrated electron flux as a function of distance along the loop. The color-coded lines represent fluxes at different electron energies, 3, 6.1, 12.3, 24.5, 48.8, 97.4, and 294.1~keV. Note the large step at the transition region, especially at low energies. [A color version of this figure is available in the online journal.].}
     \label{e_spec}
\end{figure*} 

\begin{figure*}[!htb]
 \minipage{1.\textwidth}
  \centering
  \epsscale{1.15}
    \plotone{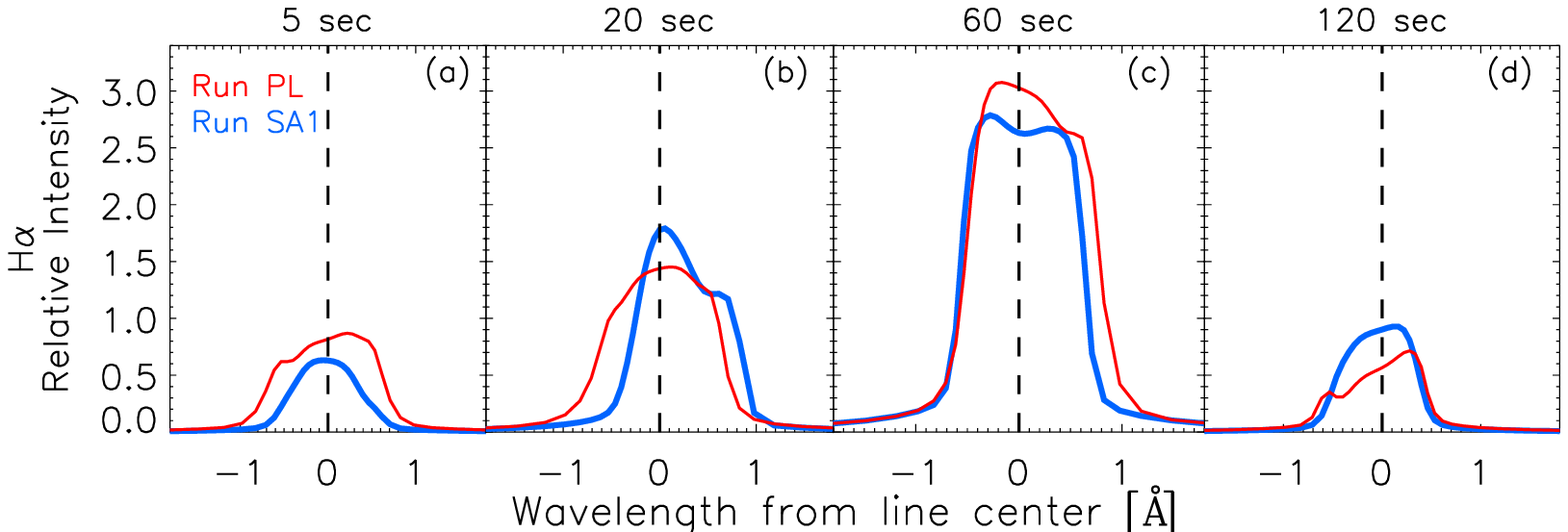}
    \caption{H$\alpha$~6563~\AA\ line profiles at $t=5$, 20, 60 and 120~s for runs PL (red line) and SA1 (blue thick line). The emission at the initial time has been subtracted and the intensity has been normalized to the continuum. The vertical dashed line indicates the line center rest wavelength.}
    \label{ha_profile_runs3_4}
\endminipage\hfill
\minipage{1.\textwidth}
  \epsscale{2.35}
  \plottwo{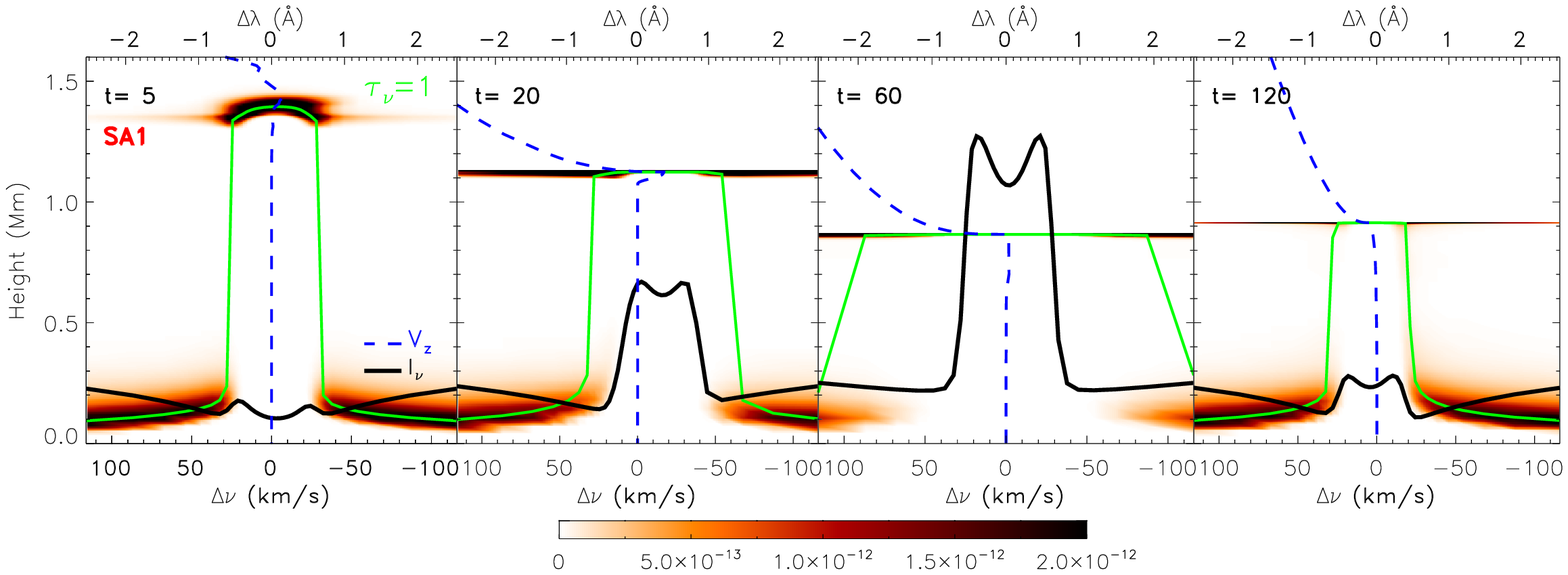}{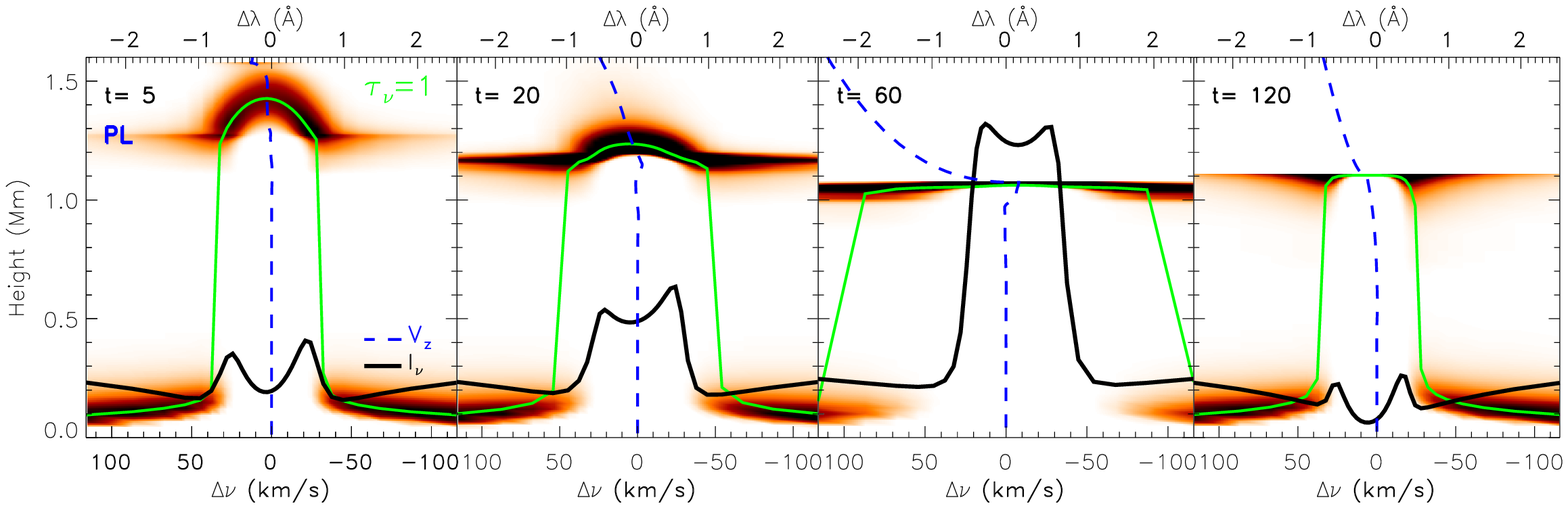}
    \caption{Intensity contribution function, as defined in equation~(\ref{eq_contribution_function}), for H$\alpha$~as a function of frequency (bottom axis) and height (vertical axes), shown as reversed color maps (darker for higher values). The frequencies are in velocity units such that positive and negative velocities represent upflows and downflows of plasma, respectively. The dashed blue line represents the atmospheric velocity stratification; the solid green line the height at which $\tau_{\nu}=1$ and the black thick line shows the line profile (including the quiet Sun emission). Top panel: SA1~Run; bottom panel: PL~Run. All panels have the same color scale as indicated on the colorbar. [A color version of this figure is available in the online journal.].}
    \label{ha_ci}
\endminipage
\end{figure*}

Such distinct temperature distributions of the plasma velocity, especially early during a flare, are manifested in the Doppler shifts of emission lines shown in Figures~\ref{ha_profile_runs3_4}, \ref{he_profile_runs3_4}, \ref{ca_profile_runs3_4} and \ref{si_profile_runs3_4}, which will be discussed later, and demonstrates their sensitive dependence on the spectra of accelerated electrons. When compared with observations, such distributions can be used as diagnostics to constrain particle acceleration mechanisms. For example, \citet{2009ApJ...699..968M} used the EUV Imaging Spectrometer (EIS) onboard {\it Hinode} to measure the Doppler velocities of emission lines at formation temperatures ranging from 0.05 to 16~MK. They found a temperature distribution of Doppler velocity akin to that of SA1 shown in Figure~\ref{vel_temp}(b). Specifically, their $-60$ to $-30$~km~s$^{-1}$ downflow speeds within 0.6--1.5~MK are comparable to those in our SA1 case at  $t=9$--12~s. They found slightly higher division temperature of $\sim$2~MK between up- and downflows. This minor difference could be partly due to the greater electron energy flux of $5 \times 10^{10}$ erg cm$^{-2}$ s$^{-1}$ estimated in their C-class flare, about five times that adopted in our simulation.

    \subsection{Evolution of the electron heating rate} \label{e-heating}
The height distribution of the electron heating rate $Q_e(s)$ is shown in the bottom row of Figure~\ref{evol_atm} at $t=5$, $20$, $60$ and $120$~s for PL (red line) and SA1 (blue thick line). In general, both runs present a peak in the upper chromosphere. However, $Q_e$ in PL is higher in the chromosphere but lower in the corona than that in SA1.

In PL the electron energy is mostly deposited in the upper chromosphere (at $z=1.18$~Mm for $t=1$~s). The electrons interact with the plasma, ionizing hydrogen and helium. A step in the heating rate appears at the chromospheric evaporation front due to the density jump (e.g., at $z \approx 3.7$~Mm in Figure~\ref{evol_atm}(n)). During the decay phase of the flare, the heating rate decreases throughout the loop, but there is more fractional reduction in the chromosphere than in the corona.

In the case of SA1, $Q_e$ is mostly concentrated around $z$ $\approx$~1.45~Mm after $t=1$~s, covering a broader region than in PL. This is because the stochastically-accelerated electrons have a quasi-thermal spectral component plus a non-thermal tail covering a broader energy range, as shown in Figure~\ref{e_spectra_top}. The quasi-thermal component carries considerable energy contents and causes significant heating in the corona that reaches temperatures of 11~MK at $t=10$~s and 23~MK at 60~s. The secondary heating peak in the lower chromosphere is due to the non-thermal tail that exceeds the power-law spectrum from $\approx$800~keV to 6~MeV. The sharp spike in the photosphere is due to the increase of the electron density at that height.

The above behaviors of the electron heating rate can be better understood by examining the energy and spatial distributions of the non-thermal electrons.
Figure~\ref{e_spec}(a) shows the energy spectra of $E^2 F(E,\,s)$ at selected distances $s$ from the top of the loop for Run~SA1 at $t=60$~s, where $F(E,\,s)= \int_{-1}^{1} v\, f(E,\, \mu, \,s) \, d \mu$ is the pitch-angle integrated electron flux. As expected, low-energy electrons suffer most of the losses at higher altitudes than the high energy electrons. This is mainly because of the $1/v$ dependence of the Coulomb collision energy-loss rate. Note that the large decrease of electron flux from $s=9$ to 10~Mm results from the change of location from above to below the transition region with a sudden increase of the atmospheric density.

Figure~\ref{e_spec}(b) shows the spatial distribution of the electron flux $F(E,\,z)$ along the loop at selected electron energies for Run~SA1 at $t=60$~s corresponding to Figure~\ref{e_spec}(a). In general, the electron flux decreases with distance from the injection site at $z=10.40$~Mm. The slope ($dF(E,z)/dz$) is steeper at lower energies because low-energy electrons loose energy faster than high-energy electrons. 
This slope also depends on the ambient density $n_e$, because $dF(E,z)/dz=n_e \big[{dF(E,N)}/{dN} \big]$, where $dF(E,N)/dN$ is generally a smooth function of column density $N$ \citep{1990ApJ...359..524M}. As such, a sharp drop in the electron flux, more prominent at lower energies, occurs at the transition region ($z=0.8$~Mm) due to the sharp ambient density rise. In addition, the slope is proportional to the electron heating rate (see equations~(10) and (11) in Paper~I) and thus accounts for its shape, especially the narrow peak at the transition region, as shown in Figure~\ref{evol_atm}(m-p).

  \section{Line emission}\label{sect:lines_emission}
One of the advantages of this work over previous studies \citep[e.g.,][]{2009ApJ...702.1553L} is the inclusion of the detailed radiative transfer calculation of emission lines treated in non-LTE and thus the capability of synthesizing line emission from simulation results at different height formations with the aim of studying how the energy deposition affects the lower atmosphere. 
In this section we will examine four emission optically thick lines, H$\alpha$, \ion{Ca}{2}~K, \ion{He}{2}~304~\AA, \ion{Si}{4}~1393~\AA, which are common observables and cover formation heights from the upper photosphere to the transition region. For example, the H$\alpha$~and \ion{Ca}{2}~K lines are routinely observed by ground-based facilities, e.g., at the Big Bear Solar Observatory \citep[BBSO, ][]{1995SoPh..161..201J, 1998SoPh..177..265J} and the Kanzelh{\"o}he Observatory \citep[KSO, ][]{2013EGUGA..15.1459P}; the Extreme-ultraviolet Variability Experiment ({\it EVE}) on board the {\it Solar Dynamics Observatory} ({\it SDO}) \citep{2012SoPh..275..115W} covers the \ion{He}{2}~304~\AA\ and \ion{Si}{4}~1393~\AA\ lines; the latter is also covered by the {\it Interface Region Imaging Spectrograph} ({\it IRIS}) \citep{2014SoPh..289.2733D}. We intend to perform detailed comparison with observations in future investigations.

In order to better understand the behavior of these lines, we write the formal solution of the radiative transfer equation for emergent intensity:
\begin{equation}
I_{\nu}^{0} =  {1\over\mu}\int_z S_\nu e^{-{\tau_\nu}\over\mu} \chi_\nu \; dz = \frac{1}{\mu} \int_{z} C_i \; dz \,\,,
\label{eq_contribution_function}
\end{equation}
where $S_{\nu}$ is the source function, which is defined as the ratio between the emissivity to the opacity of the atmosphere; $\tau_{\nu}$ is the monochromatic optical depth and the integrand $C_i$ is the so-called intensity contribution function, which represents the intensity emanating from height $z$.


    \subsection{H$\alpha$~6563~\AA\ line emission}
The H$\alpha$~line is one of the most commonly observed lines that allow us to study the chromospheric response during a solar flare. It is sensitive to the flux of non-thermal electrons precipitating to the chromosphere \citep{1976saop.book..141S, 2009A&A...499..923K, 2015ApJ...804...56R} and a complex line formed in a broad height range, from the upper photosphere to the lower chromosphere, making its interpretation non-trivial.

The top panels of Figure~\ref{ha_profile_runs3_4} shows the temporal evolution of the resulting excess H$\alpha$~line profile, where the quiet-Sun emission at $t=0$~s has been subtracted to emphasize the changes during the flare and the differences between Runs PL (red line) and SA1 (blue line). In both runs, the line core presents a flattening due to the sudden behavior change of the source function in a very thin atmospheric layer, as previously reported by \citet{2015ApJ...804...56R}. The line is broader and the intensity at the core is stronger for PL most of the time, while SA1 shows stronger asymmetry (mostly blueshifts associated with upflows) due to the higher velocities in the chromosphere. Close to $t_{max}=60$~s the profile presents a stronger blueshifted peak in the line center and redshifts in the wings, indicating plasma upflows (evaporation) from the chromosphere to the transition region and downflows in the chromosphere. 

\begin{figure}[tbh]
\centering
  \epsscale{1.23}
  \plotone{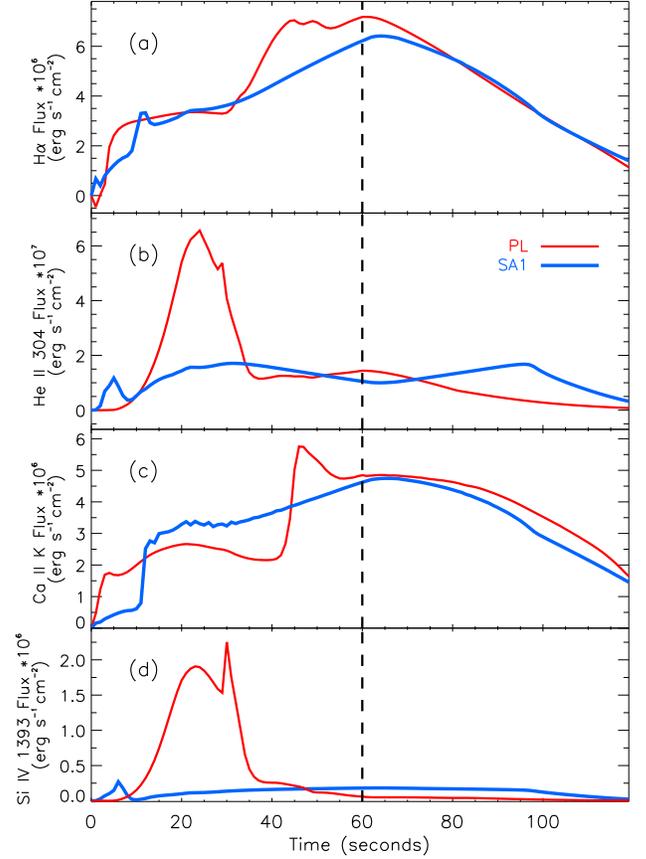}
  \caption{H$\alpha$, \ion{He}{2}~304~\AA, \ion{Ca}{2}~K and \ion{Si}{4} light curves. The emission at the initial time has been subtracted. The dashed line indicates $t_{max}$. [A color version of this figure is available in the online journal.].}
     \label{ha_he_ca_si_light_curves}
\end{figure}

The above behavior can be better understood by examining the contribution function, as shown in Figure~\ref{ha_ci}. We find that the photons in the wings of the line originate from the low chromosphere (at $z=0.12$~Mm), where the plasma velocity is almost zero and almost constant in time. Therefore the wing emission changes little in time and is very similar in both runs. In contrast, the line core is formed in the transition region, whose height changes in time and where the plasma velocity suddenly increases due to upflows. For SA1, this region is situated at a lower height and thus the overall height range of H$\alpha$~line formation is narrower than that of PL. The PL~model has lower velocities and covers a broader formation height range at the line core. Moreover, the plasma temperature and density play an important role in the formation of the line and might explain the differences in the core emission between the two runs \citep{2012ApJ...749..136L}. 

Integrating the intensity along the H$\alpha$~line profile for a width of $\Delta \lambda=$3.8~\AA\ yields the light curve shown in Figure~\ref{ha_he_ca_si_light_curves}(a). As can be seen, for SA1 the H$\alpha$~emission is directly correlated with the evolution of the flux of electrons, peaking only 4~s after $t_{max}$. However, for PL, the emission exhibits a plateau after 47~s, peaking exactly at $t_{max}$. 

    \subsection{\ion{He}{2}~304~\AA\ line emission}

\begin{figure*}[!htb]
  \minipage{1.\textwidth}
  \centering
  \epsscale{1.15}
    \plotone{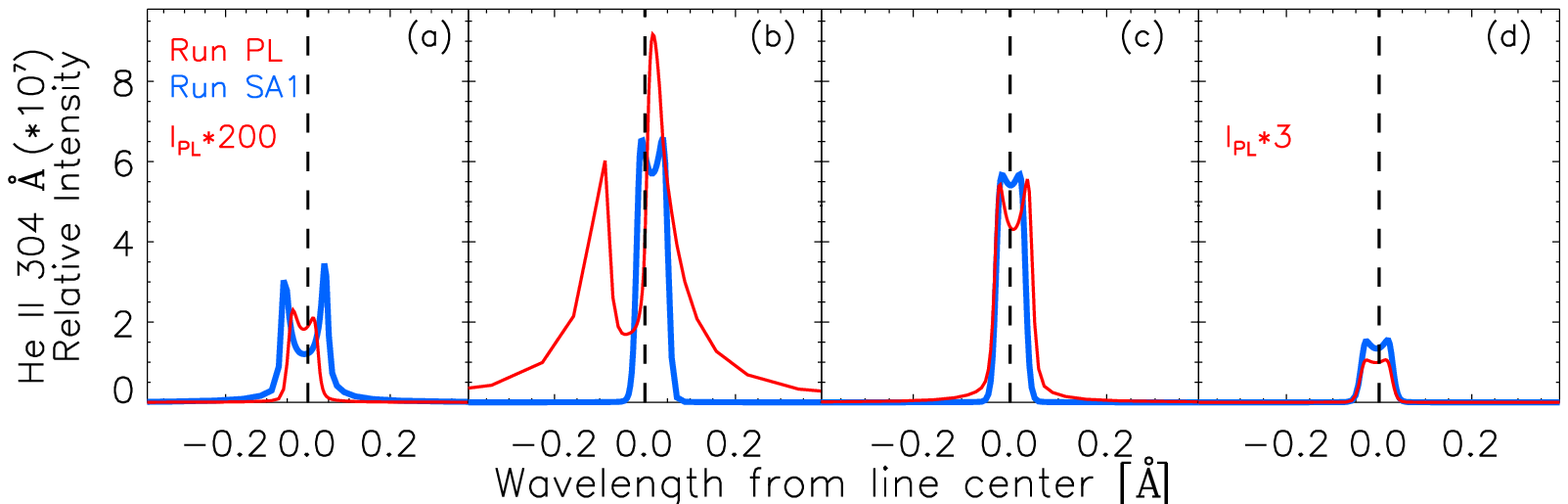}
    \caption{\ion{He}{2}~304~\AA\ line profiles at $t=5$, 20, 60 and 120~s for runs PL (red line) and SA1 (blue thick line), similarly as Figure~\ref{ha_profile_runs3_4}.}
    \label{he_profile_runs3_4}
\endminipage\hfill
\minipage{1.\textwidth}
  \epsscale{2.35}
  \plottwo{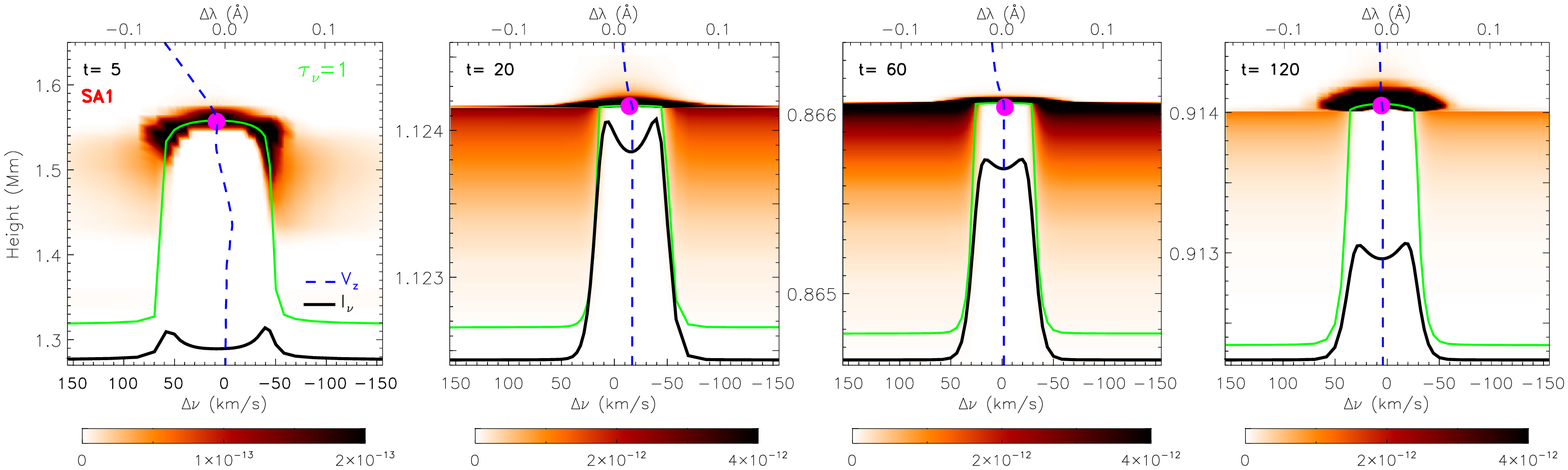}{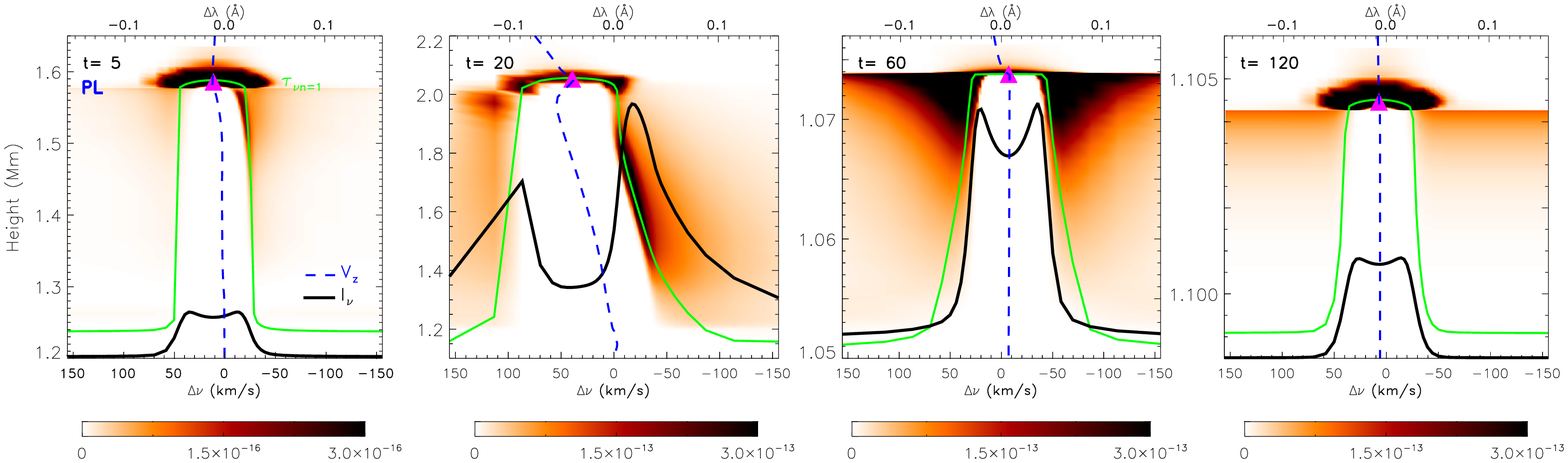}
    \caption{Intensity contribution function for \ion{He}{2}~304~\AA\ as a function of frequency and height, similarly as Figure~\ref{ha_ci}. Left panel: SA1~Run; right panel: PL~Run. The pink circle (in the SA1 case -- top) and triangle (in the PL case -- bottom) mark the average formation height of the line center. The color scale of each panel is shown in its respective colorbar. Note that the line profiles are not in scale. [A color version of this figure is available in the online journal.].}
    \label{he_ci}
\endminipage
\end{figure*}

The \ion{He}{2}~304~\AA\ line is an optically thick line associated with the transition $1s \; ^2S_{1/2}-2p \; ^2P^0_2$. Under ionization equilibrium conditions, in general, it is formed in a narrow layer in the transition region, corresponding to temperatures of $\approx 5-8 \times 10^4$ K \citep{2010A&A...521A..21O}. However, \citet{2014ApJ...784...30G} noted that the details of its formation is still not well understood. \citet{1988assu.book.....Z} showed that the photoionization and recombination processes play an important role in the \ion{He}{2}~304~\AA\ emission. \citet{1975MNRAS.170..429J} (see also \citet{2003A&A...400..737A}) found that the observed intensity of the helium-lines is higher than that estimated from observations of other EUV lines. They noted that the inclusion of high-energy electrons and cold ions in the models enhances the synthetic intensities. Based on this idea, \citet{1992ApJ...386..364L} modeled the \ion{He}{1} and \ion{He}{2} emission during the impulsive phase of a flare by including the effects of high-energy electrons. 

In Figure~\ref{he_profile_runs3_4} we note large differences between PL and SA1, indicating that the \ion{He}{2}~304~\AA\ emission is very sensitive to the spectra of the non-thermal electrons. In fact, the intensity in PL changes by two orders of magnitude within 9~s of the simulation. Whenever there are large velocity variations in the formation region, the line profile shows strong asymmetries. In particular, the line profile of the PL model at $t=20$~s shows a strong blueshift due to the evaporation of material to upper layers.

\begin{figure}[tbh]
\centering
  \epsscale{2.5}
  \plottwo{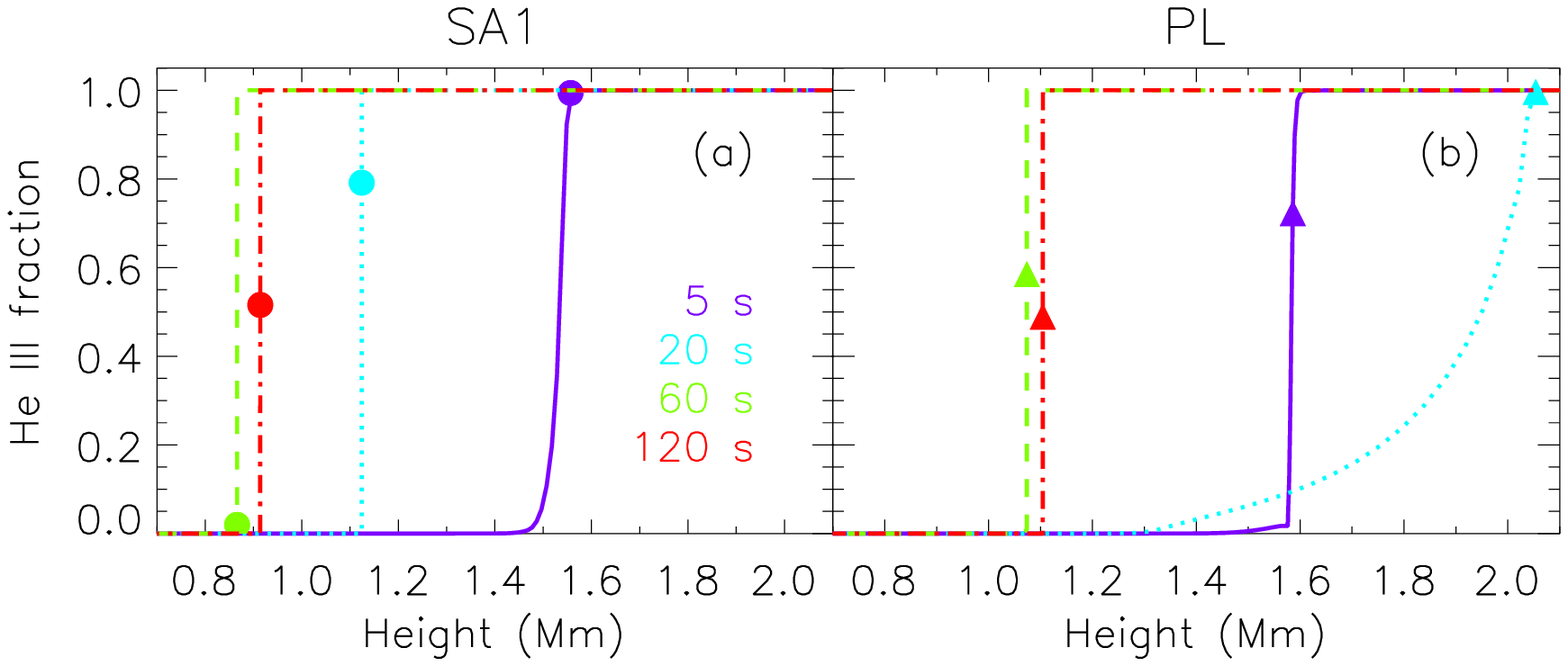}{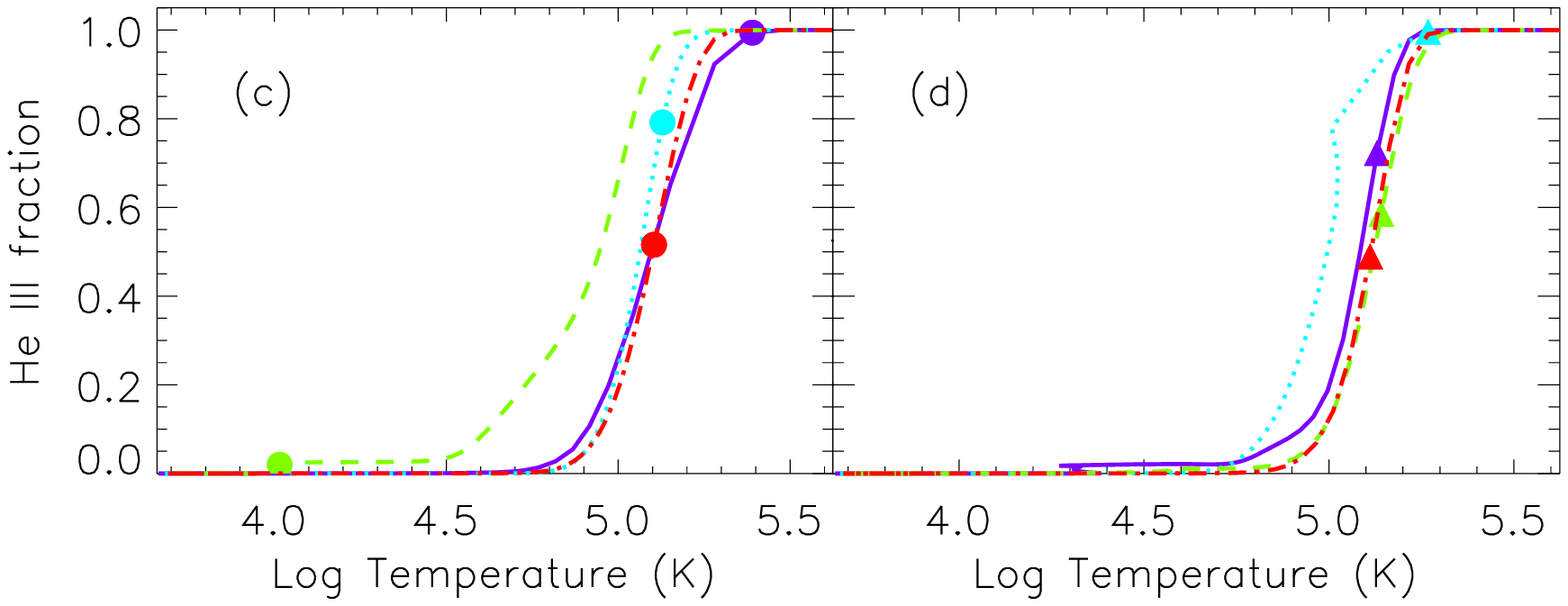}
  \caption{\ion{He}{3} ionization fraction as a function of height (top) and temperature (bottom) for Runs~SA1 (left) and PL (right) at $t=5$, 20, 60 and 120~s. The pink circles (in the SA1 case -- left) and triangle (in the PL case -- right) mark the height and temperature formation of the line center as shown in Figure~\ref{he_ci}. [A color version of this figure is available in the online journal.].}
   \label{he_ioniz_frac}
\end{figure}

Studying the monochromatic optical depth (green line in Figure~\ref{he_ci}), we find that the photons of the \ion{He}{2}~304~\AA\ line are emitted in a very narrow region situated at the bottom of the transition region. The height range of the formation region varies between 5 and 300~km for the PL model, being even narrower (between 2 and 5~km after $t=10$~s) for the SA1 model. Figure~\ref{he_ioniz_frac} shows the \ion{He}{3} ionization fraction as a function of height and temperature. It indicates that the \ion{He}{2} atoms become completely ionized within a narrow height range in the transition region. At $t=20$~s, the PL model presents a gradual change of the ionization fraction over a broad height range, which is related to the broad formation heights shown in Figure~\ref{he_ci} at this time. 

From the intensity contribution function of Figure~\ref{he_ci} we find that most of the intensity comes from $\approx \pm 0.02$~\AA\ from the line center, and thus the line center appears as a dip. The line profile is very sensitive to plasma velocity changes as shown, e.g. at $t=20$~s for the PL model. The temperature range at the formation height is of the order of 1--2$\times 10^5$ K (see the pink symbols in Figure~\ref{he_ioniz_frac}), more than twice higher than the expected values from \citet{2010A&A...521A..21O} and consistent with the simulation results reported by \citet{2014ApJ...784...30G}. In addition, \citet{1993ApJ...406..346J} found that the \ion{He}{2}~304~\AA\ emission is formed by collisional excitation in the quiet sun, but by the photoionization-recombination process in active regions and during flares. This is consistent with the high temperature values in the formation region revealed in our simulation.

We note in passing that, as shown in Figure~\ref{ha_he_ca_si_light_curves}, the integrated intensity for the PL model experiences a rapid increase at $t=11$~s and peaks at $23$~s, which coincides with an increase in electron density at transition region temperatures. In contrast, such a peak is absent in the light curve of the SA1 model, which exhibits more mild temporal variations.

    \subsection{\ion{Ca}{2}~K 3934~\AA\ line emission}

\begin{figure*}[!htb]
  \minipage{1.\textwidth}
  \centering
  \epsscale{1.15}
    \plotone{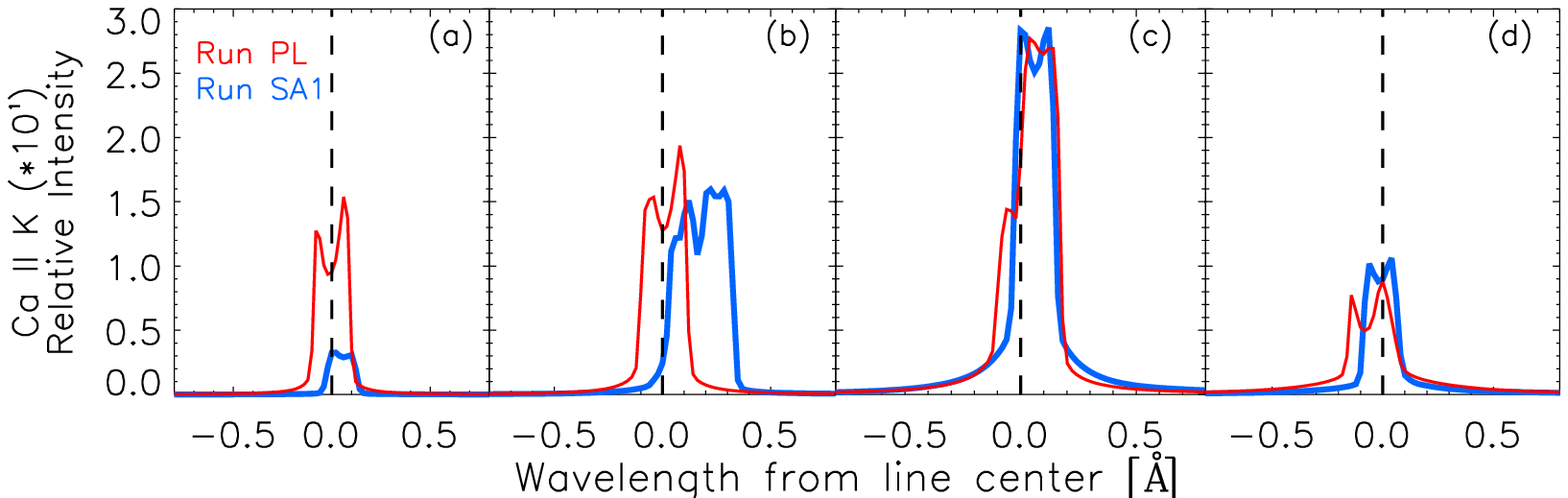}
    \caption{\ion{Ca}{2}~K~3934~\AA\ line profiles at $t=5$, 20, 60 and 120~s for runs PL (red line) and SA1 (blue thick line), similarly as Figure~\ref{ha_profile_runs3_4}.}
    \label{ca_profile_runs3_4}
\endminipage\hfill
\minipage{1.\textwidth}
  \epsscale{2.35}
  \plottwo{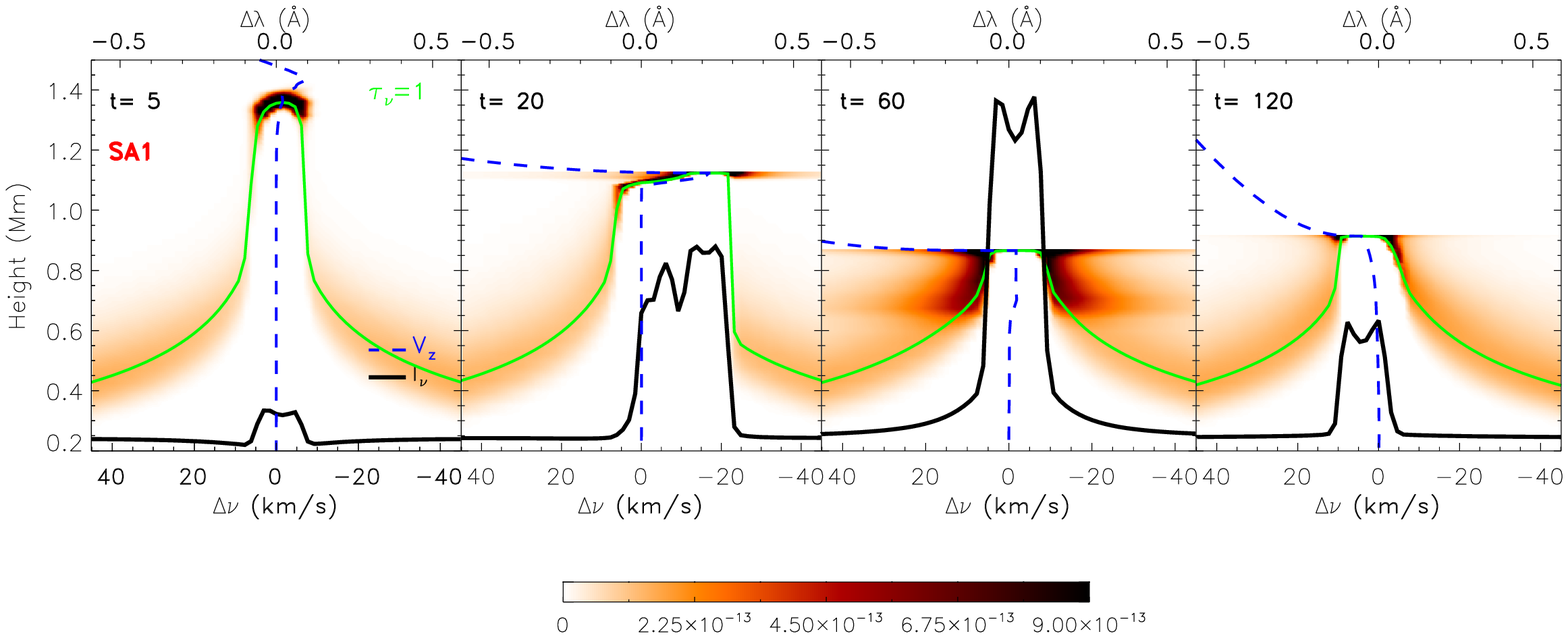}{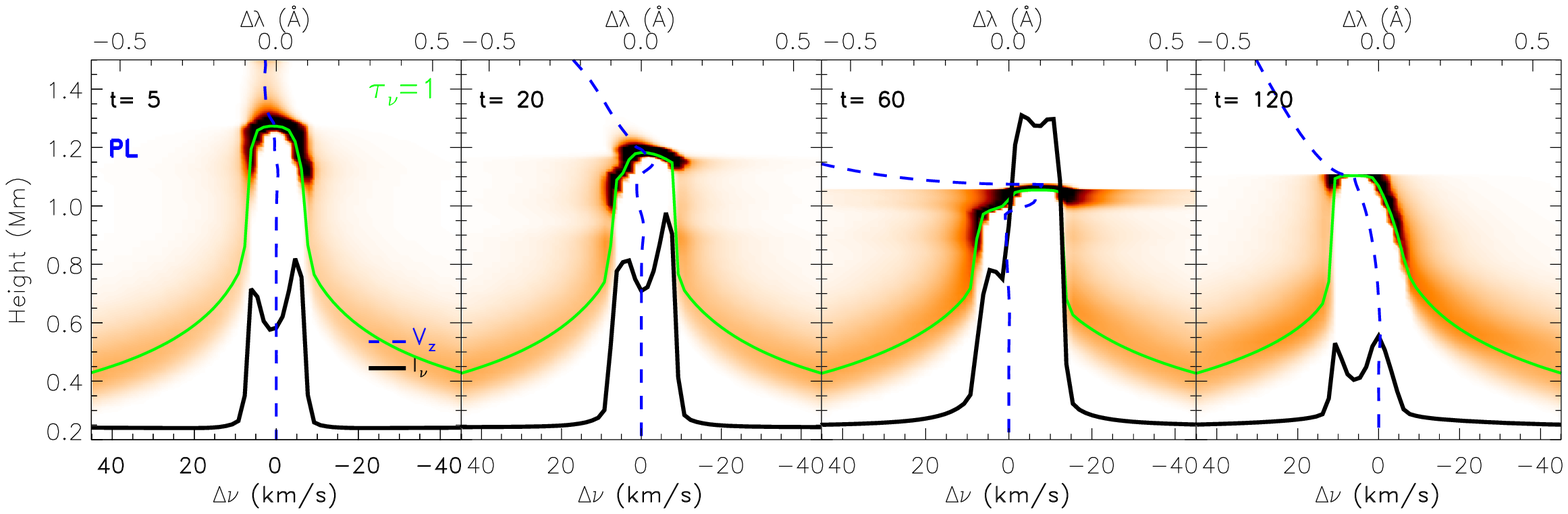}
    \caption{Intensity contribution function for \ion{Ca}{2}~K as a function of frequency and height, similarly as Figure~\ref{ha_ci}. Top panels: SA1~Run; bottom panels: PL~Run. All panels have the same color scale as indicated on the colorbar. [A color version of this figure is available in the online journal.].}
    \label{ca_ci}
\endminipage
\end{figure*}

The \ion{Ca}{2}~K line is formed in the transition $2s - 2p_1$, absorbing photons at 3934~\AA. \citet{1982ApJ...253..330V} modeled the conditions at which the line profile is formed, assuming complete redistribution (CRD), requiring low electron densities ($2~\times~10^{10}$~cm$^{-3}$) and a relatively low ionization degree of hydrogen. \citet{1993A&A...274..571P} improved the results by considering the partial frequency redistribution approximation (PRD) in the model, getting higher intensity values. To model the \ion{Ca}{2}~K line profile it is important to keep in mind that the wings are formed under LTE conditions, while the core is formed under non-LTE conditions \citep{2006A&A...449.1209L}. Another point to take into account in the modeling of this line is the formation height during the quiet Sun, which ranges from the chromosphere above the minimum temperature region up to the transition region.

The \ion{Ca}{2}~K line profile (Figure~\ref{ca_profile_runs3_4}) shows strong asymmetries due to the large plasma velocities at these heights, consistent with that observed by \citet{1993AdSpR..13..311C} during the early phase of a solar flare. SA1 presents stronger redshifted profiles (specially at $t=20$~s) due to its higher velocities than PL, indicating that the \ion{Ca}{2}~K emission is sensitive to plasma velocities. 

The monochromatic optical depth (green line in Figure \ref{ca_ci}) for \ion{Ca}{2}~K shows that the wings of the line are formed at $z=0.41$~Mm in both runs, not changing in time. Instead the core is formed in the transition region, which is located at a lower height for the SA1 model (e.g., see~Figure~\ref{evol_atm}). As shown in Figure~\ref{ca_ci}, the contribution function is weaker in the wings than in the line core, where the velocity changes rapidly with height. In fact, the plasma in the formation region is subject to downflows and upflows, which is manifested as the distortion in the line profile near the core. Nevertheless the wings still exhibit a symmetric shape. We also noted that the \ion{Ca}{2}~K line wings are formed higher in the chromosphere than the H$\alpha$~line for both runs.

As Figure~\ref{ha_he_ca_si_light_curves}(c) shows, the \ion{Ca}{2}~K intensity in the PL model responds faster to the non-thermal electrons than in SA1, reaching higher intensity values at initial times; afterwards it follows a similar behavior as the flux of the electrons, $\mathfrak{F}$. The flux for the SA1 model increases almost gradually, starting at $t=12$~s up to $t=65$~s and decreases monotonically afterwards.

    \subsection{\ion{Si}{4}~1393~\AA\ line emission}\label{sect:Si_line_emission}

\begin{figure*}[!htb]
  \minipage{1.\textwidth}
  \centering
  \epsscale{1.15}
    \plotone{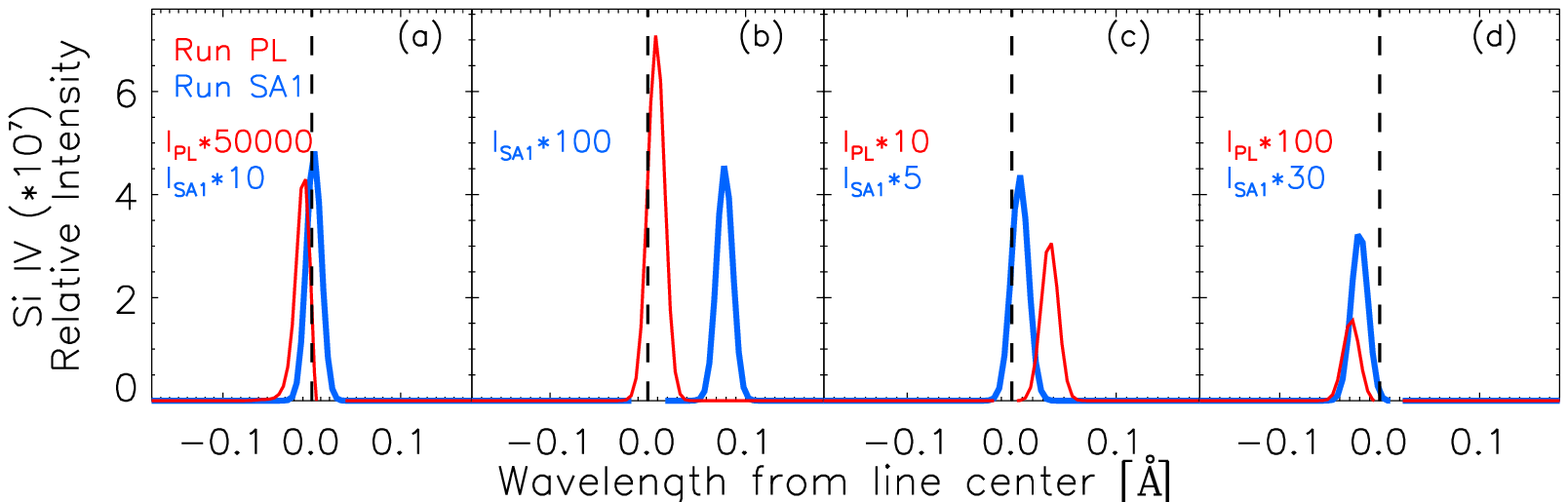}
    \caption{\ion{Si}{4}~1393~\AA\ line profiles at $t=5$, 20, 60 and 120~s for runs PL (red line) and SA1 (blue thick line), similarly as Figure~\ref{ha_profile_runs3_4}. [A color version of this figure is available in the online journal.].}
    \label{si_profile_runs3_4}
\endminipage
\end{figure*}

The \ion{Si}{4}~1393~\AA\ line is an optically-thin line formed in a narrow region located in the upper chromosphere, which under ionization equilibrium conditions corresponds to a temperature range around $(8 \pm 2) \times 10^4$ K. \citet{2015ApJ...799L..12D} showed that the correlation between the non-thermal line broadening and the intensity is reproduced with simulations only when the non-equilibrium ionization is taken into account. This is because the non-equilibrium ionization leads to the presence of Si$^{3+}$ ions over a much wider range of temperatures than under ionization equilibrium \citep[see e.g.][]{2013ApJ...767...43O}.

We synthesized the intensity profile by using the CHIANTI abundances \citep{1997A&AS..125..149D, 2013ApJ...763...86L} and the plasma properties resulting from our atmosphere. The bottom panels of Figure~\ref{si_profile_runs3_4} show the evolution of the \ion{Si}{4} line profile. This evolution can be described as an initial rapid blueshifted excursion due to chromospheric evaporation of material to the corona, followed by chromospheric compression causing redshifts and then blueshifts again in the late phase. 

Comparing the two runs, we note that, in general, PL presents larger blueshifts, while SA1 has greater redshifts, especially early in the simulation. This is due to their different plasma velocity distributions around the \ion{Si}{4} line formation temperature shown in Figures~\ref{vel_temp}(a) and (b). This trend is qualitatively consistent with the recent simulation result of \citet{2014Sci...346B.315T} which shows that non-thermal electrons are required to produce blueshifted \ion{Si}{4} emission, while conductive heating alone tends to produce only redshifted emission. The latter case resembles the strong heating in the corona rather than in the lower atmosphere in our Run~SA1, due to the presence of its prominent quasi-thermal component of electrons.
We also note differences in the large temporal variations of the intensity between the two runs by up to orders of magnitude, the largest among all four emission lines studied here, as shown in Figures~\ref{si_profile_runs3_4} and \ref{ha_he_ca_si_light_curves}. Early in the simulation during $t=$20 and 30~s, PL presents a broad peak in the integrated intensity, more than 7 times higher than that of SA1 (see Figure~\ref{ha_he_ca_si_light_curves}(d)). This is due to its larger electron density hump in the transition region, as shown in Figure~\ref{evol_atm}, and thus greater emission measure.

    \section{Summary and discussions}\label{Sect:conclusion}
The aim of this paper is to investigate the response of the solar atmosphere to the energy input by accelerated electrons. To achieve this, we have extended our earlier study (paper I) by including radiative transfer calculations. Specifically, we combined the Stanford Unified Acceleration-Transport code with the radiative hydrodynamics RADYN code. Our primary focus is to compare the results from the more realistic stochastic acceleration model with that from an {\it ad hoc} power-law injection model and to obtain synthetic line emissions from our simulation results, which can provide new constraints for the particle acceleration mechanism.
Our main findings are the following:

\begin{enumerate}

\item In general, the temporal evolution of the atmosphere is determined not only by the energy flux but also the spectral shape of non-thermal electrons.
The results of our Runs~PL (power-law injection) and SA1 (stochastic acceleration) are in qualitative agreement with those of Runs~O and N of Paper~I \citep{2009ApJ...702.1553L}, respectively. Stochastically accelerated electrons lead to stronger chromospheric evaporation with higher coronal temperatures and plasma velocities (see Table~\ref{table_runs}) because of their prominent quasi-thermal spectral component. Hydrodynamic shocks form in both cases, lasting longer in PL, but appearing earlier in SA1.

\item The spatial distribution of the electron energy deposition rate per unit volume is concentrated in the upper chromosphere for both runs (see bottom row of Figure~\ref{evol_atm}). In terms of the electron heating rate per unit atmospheric mass, most of the directly deposited energy is radiated away in the chromosphere for Run~PL (Figure~\ref{energy_terms_balance_run1}).
However, due to the quasi-thermal spectral component, Run~SA1 has a significant amount of energy per unit mass directly deposited in the corona, rather than in the chromosphere (Figure~\ref{energy_terms_balance_run2}).
Downward thermal conduction then carries the bulk of this energy to heat the transition region, where radiative loss is not as efficient as in the chromosphere. This is why chromospheric evaporation is much stronger in SA1 than in PL, despite their identical total electron energy flux.

\item In both cases, most of the energy exchange occurs in the lower atmosphere, and most of the net energy gain of the plasma is in the form of thermal energy, which dominates over ionization energy expenses. The thermal energy increase is about two orders of magnitude higher for SA1 than for PL.

\item For different SA runs of different peak electron energy flux $\mathfrak{F}_{max}$, the maximum upflow velocity increases almost linearly with the $\mathfrak{F}_{max}$, while the maximum downflow velocity increases at a lower rate (see Appendix~\ref{sect:results_flux}). The temperature peaks at $t_{max}$ is independent of $\mathfrak{F}_{max}$.

\item Surprisingly, explosive chromospheric evaporation with upflow speeds of hundreds of km~s$^{-1}$ is present in our three SA runs with wide-ranging electron energy fluxes $\mathfrak{F}_{max}$ from 10$^8$ to 10$^{10}$~erg~s$^{-1}$~cm$^{-2}$. In contrast, explosive evaporation only occurred when $\mathfrak{F}_{max}~> 3~\times~10^{10}$~erg~s$^{-1}$~cm$^{-2}$ in early simulations by \citet{1985ApJ...289..414F}, who used a power-law injection with a spectral index of 4 and low-energy cutoff of 20~keV. Such energy flux was
widely quoted as the sole quantity that determines gentle vs. explosive evaporation. Our result demonstrates that the shape of the electron spectrum is at least equally, if not more, important as the total energy flux, for the reasons noted above. This is consistent with the conclusions independently reached by \citet{2015arXiv150608115R} and \citet{2015ApJ...809..104A}.
\end{enumerate}

In order to study how the different models affect to the energy deposition in the lower chromosphere, we obtained the emission in several wavelength ranges formed at different heights from the upper photosphere to the transition region. According to the line emission for both runs, we find that the emission in H$\alpha$~and \ion{Ca}{2}~K (two lines that cover the whole chromosphere) seems to be more dependent on the electron flux variation than emission in other wavelengths, following a similar triangular temporal profile as the flux of electrons, $\mathfrak{F}$.
The characteristics of the specific emission lines are as follows:

\begin{enumerate}

\item The H$\alpha$~line is a broad line covering the chromosphere, from 0.12~Mm up to the transition region, where the temperature changes drastically. The SA1~Run has higher chromospheric velocities and therefore stronger line-profile asymmetries than the PL~Run. The core formation is located at a height where the plasma velocity changes drastically (from almost zero in the lower chromosphere to more than 50~km~s$^{-1}$), covering a broader region for the PL~Run.
Close to $t_{max}$ the profile presents a stronger blueshifted peak in the line center and redshifts in the wings, indicating plasma upflows (evaporation) from the chromosphere to the transition region and downflows in the chromosphere.
The line intensity is temporally correlated with the electron energy flux, peaking almost at its apex $t_{max}$. 

\item Our synthetic \ion{He}{2}~304~\AA\ line is formed within a temperature range from $1\times 10^4$ to $2.5\times 10^5$~K, which is broader than previously reported for the quiet Sun, when this line is formed by collisional excitation.
However, in agreement with the results from \citet{1993ApJ...406..346J} we find that, for active regions and during flares, the photoionization-recombination processes are more important.
Thus the high temperatures at the formation height in our simulation are not unexpected.
This line is formed at the bottom of the transition region, covering a height range between 5 and 300~km for the PL model, being even narrower for the SA1 model. The shape of the line profile is strongly affected by the plasma velocity at the formation height, showing asymmetries at locations where large velocity gradients are present.
The integrated intensity is strongly correlated with the electron density, as manifested in its large enhancement from 11 to 36~s for the PL model (see Figure~\ref{ha_he_ca_si_light_curves}(b)).

\item The \ion{Ca}{2}~K line is formed in the chromosphere above the temperature minimum, at 0.41~Mm up to the transition region, where the temperature changes drastically. The line profile shows strong asymmetries, consistent with the observations of \citet{1993AdSpR..13..311C} during the early phase of a solar flare. 
The \ion{Ca}{2}~K flux variation responds faster to the non-thermal electron heating in PL than in SA1.

\item The \ion{Si}{4} line is optically thin, formed in the upper chromosphere. Run~PL presents stronger chromospheric evaporation, greater up- and downflow velocities and higher electron density values, resulting in seven times larger \ion{Si}{4} intensity than run~SA1.

\end{enumerate}

Our results have demonstrated that the emission line profiles and light curves of the integrated intensities can provide useful plasma diagnostics as well as constraints for particle acceleration models. For example, the PL run exhibits similar behaviors of large increases in the light curves of both transition region lines, \ion{He}{2}~304~\AA\ and \ion{Si}{4}~1393~\AA, early during the simulation (see Figure~\ref{ha_he_ca_si_light_curves}). Such large increases are in contrast with the rather slow rise and even drop at certain times in the SA1 case. Such predicted distinction in the temporal evolution can be checked against observations to shed light on the shape of non-thermal electrons.

Our results also have important implications for sunquakes during solar flares \citep{1998Natur.393..317K}. Such seismic signals are believed to be caused by the response of the lower atmosphere to impulsive heating by high-energy particles \citep{2007ApJ...670L..65K, 2008SoPh..251..641Z}, among other proposed mechanisms \citep[][]{2008SoPh..251..627L, 2011SSRv..158..451D, 2012SoPh..277...59F}. Specifically, sunquakes can be produced by hydrodynamic shocks propagating downward from the chromosphere to the photosphere at onsets of flares. In our simulations, we find that the hydrodynamic response to collisional heating results in initial downflows, especially in the SA1 model, at velocities up to the order of 45 km s$^{-1}$ within the first 10~s of the flare (see Figure~\ref{vel_max}(b)). Such high speed downflows are sufficient to create hydrodynamic shocks in the lower chromosphere and can be responsible for the sunquakes in the photosphere, as demonstrated in the simulations of \citet{2008SoPh..251..641Z}.

The approach of combined particle acceleration-transport and radiative hydrodynamic simulation presented in this paper has opened a new chapter in modeling solar flare dynamics. One important future improvement, among others, is to use nonparametric inversion of acceleration parameters from HXR observations \citep{2010ApJ...712L.131P, 2013ApJ...777...33C} as inputs to our model. This will allow us to more accurately and realistically simulate the atmospheric evolution as well as the chromospheric and transition region emission lines, which can be checked against high-resolution spectroscopic observations, such as those from {\it IRIS} and DST/IBIS \citep[e.g.][]{2015ApJ...804...56R}. Such comparative investigations will provide promising new constraints to the coupled processes in solar flares, including radiative transfer, hydrodynamics, and eventually particle acceleration.

\acknowledgments
This work was supported by NASA LWS grant NNX13AF79G and H-SR grant NNX14AG03G to Stanford University. The research leading to these results has received funding from the European Community's Seventh Framework Program (FP7/2007-2013) under grant agreement no. 606862 (F-CHROMA) and ERC grant agreement no. 291058 (CHROMPHYS). CHIANTI is a collaborative project involving George Mason University, the University of Michigan (USA) and the University of Cambridge (UK). We thank Joel Allred, Tiago Pereira, John Mariska, and Paul Bryans for helpful discussions.

   \appendix \label{Appendix}
\section{A.~Distribution of Energy Terms}\label{sect:internal_energy}
In this section, we examine the spatial distribution and temporal evolution of different terms of the energy equation~(\ref{eq_energy_conservation}), in order to better understand their contribution to the internal energy of the plasma.

\begin{equation}
\frac{\partial \rho \epsilon}{\partial t}+\frac{\partial \rho \upsilon \epsilon}{\partial z}+(p+q_{\upsilon})\frac{\partial \upsilon}{\partial z} + \frac{\partial}{\partial z}(F_c + F_{\rm r, \, thin} + F_{\rm r, \, detail}) - Q_{\rm init} - Q_e - Q_{\rm XEUV}=0,
\label{eq_energy_conservation}
\end{equation}

\begin{figure*}[!htb]
\centering
   \epsscale{1.15}
   \plotone{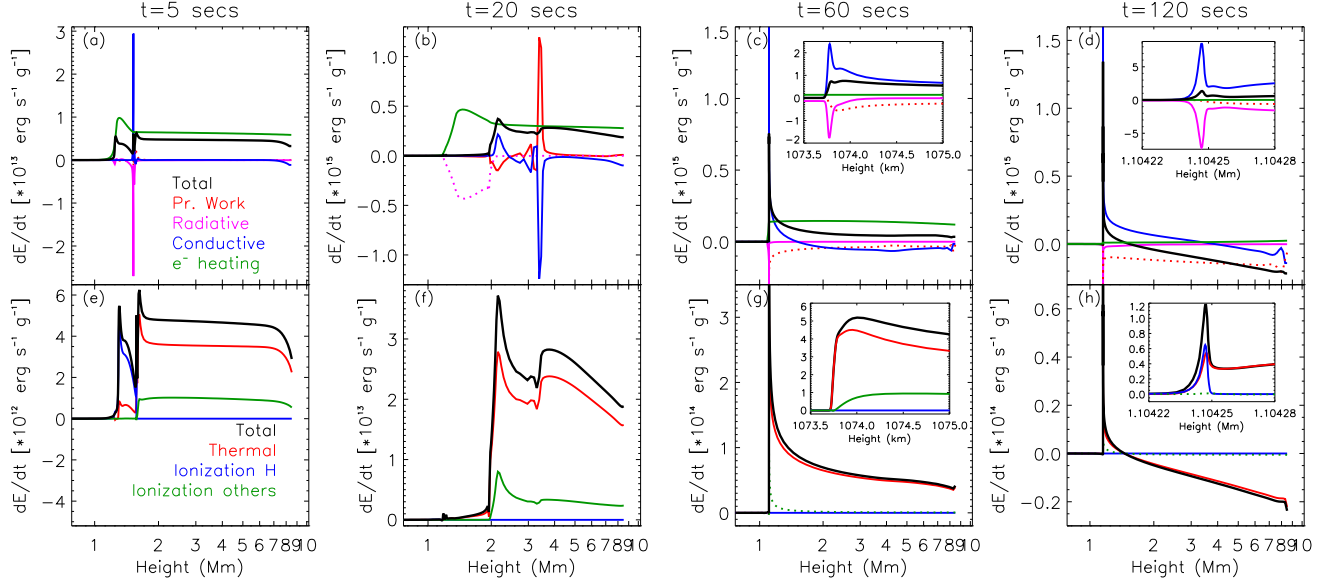}
     \caption{Variations with height of different energy terms for PL at $t=5$, 20, 60 and 120~s. Note that the X-axis is in logarithmic scale.
     Top: rate of change of major energy terms, including the total internal energy (black thick line), radiative heating (pink), conductive heating (cyan), electron heating (green), and compression work (red). 
     Bottom: rate of change of the total internal energy (black), divided into thermal energy (red), hydrogen ionization and excitation energy (blue) and ionization and excitation energy for elements other than hydrogen (green).
The insets show enlarged views near the transition region. [A color version of this figure is available in the online journal.].}
     \label{energy_terms_balance_run1}
\end{figure*}

Here $z$ is distance, $t$ time, $\rho$ density, $\upsilon$ velocity, $p$ pressure, $\epsilon$ internal energy per unit mass, $q_{\upsilon}$ the viscous stress, and $F_c$ and $F_r$ are conductive and radiative fluxes. The specific energy terms include: viscous heating $-q_{\upsilon} (\partial \upsilon / \partial z)$, compression work $-p (\partial \upsilon / \partial z)$, conductive heating $- \partial F_c/ \partial z $, the heating function $Q_{\rm init}$ to create the initial atmosphere, collisional heating $Q_e$ by non-thermal electrons, X-ray plus EUV heating $Q_{\rm XEUV}$, the thin radiative heating $- \partial F_{\rm r, \, thin} / \partial z$ of elements treated in LTE and the radiative flux $- \partial F_{\rm r, \, detail} / \partial z$ calculated from solving the radiative transfer equation. Here $F_{\rm r, \, detail}$ is integrated over wavelength and includes about 34000 wavelengths in the temperature range between 0.1 and 10~MK, covering the continua from 1 to 40000~\AA\ and spectral lines of Hydrogen, Helium, Calcium and Magnesium, treated in Non-LTE.

Since the atmospheres evolves differently in the two runs, we discuss them individually. In order to avoid complex graphics, here we show only the energy terms with significant contributions to the total internal energy. 

    \subsubsection{A1.~PL Run}\label{sect:internal_energy_run1}
The top panels of Figure \ref{energy_terms_balance_run1} show the rate of change of different energies per unit mass (cf., heating rate per unit volume in the fourth row of Figure~\ref{evol_atm}): internal energy (black thick line), radiative heating (pink), conductive heating (cyan), electron heating (green), and compression work (red). Most of the energy exchange occurs in the lower atmosphere and the main contribution to the internal energy change is the electron heating rate. At $t=5$~s and $z=1.55$~Mm (within the chromosphere), the conductive and electron heating rates are the main source of heating, which are mostly balanced by the radiative loss term, explaining the small temperature changes at this time (see Figure~\ref{evol_atm}).

At $t=20$~s, deep in the chromosphere, electron heating $Q_e$ is largely balanced by radiative losses. However, in the transition region (near $z=2$~Mm) and corona, it dominates over the loss terms and results in an increase in the internal energy and temperature. Around the chromospheric evaporation front at $z=3.7$~Mm, the pressure work due to the negative velocity gradient leads to a local temperature enhancement (see Figure~\ref{evol_atm}). Thus, compensating conductive cooling.

At later times, conductive heating in the transition region becomes important. For example, at $t=60$~s and $z=1.07$~Mm, the conductive heating rate increases in a height range of less than 150~m from almost zero to $3 \times 10^{15}$~(erg~s$^{-1}$~g$^{-1}$). This is also the region where most of the non-thermal electron energy is deposited (see bottom row of Figure~\ref{evol_atm}).

The bottom panel of Figure \ref{energy_terms_balance_run1} shows the change rate of the total internal energy (black thick line) divided into the thermal energy of the material (red), the sum of ionization energy of hydrogen (blue) and the rest of the constituent atoms (green). As can be seen, the thermal energy change dominates at all times, but the total ionization energy of atoms other than hydrogen is important at early times.

    \subsubsection{A2.~SA1 Run}\label{sect:internal_energy_run2}
\begin{figure*}[tbh]
\centering
   \epsscale{1.15}
   \plotone{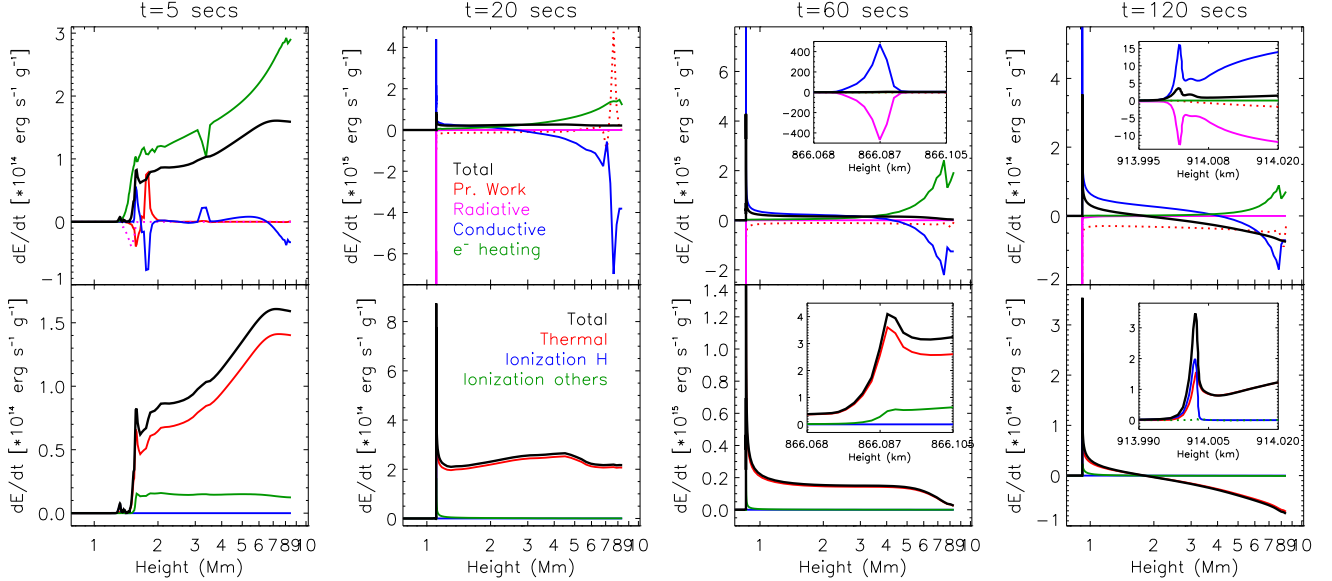}
     \caption{Same as Figure~\ref{energy_terms_balance_run1}, but for Run~SA1. Note that the X-axis is in logarithmic scale and that at $t=60$ and 120~s, the peaks at the transition region height are truncated in the main panels but are displayed in full in the insets. [A color version of this figure is available in the online journal.].}
     \label{energy_terms_balance_run2}
\end{figure*}

In the SA1 model, the electron heating rate per unit mass, the major contributor to the total internal energy increase (see Figure~\ref{energy_terms_balance_run2}), is primarily located in the corona, rather than the transition region or chromosphere as in the case of the PL model. This results in the significant increase of the coronal temperature (see Figure~\ref{evol_atm}). At $t=5$~s the pressure work balances the conductive cooling at $z=1.86$~Mm (above the transition region), where the electron density is higher.

At $t=20$~s the bulk of pressure work has moved to the upper corona and is still balanced by conductive cooling. At $z=1.125$~Mm, where the transition region is located, the conductive heating rate increases sharply up to $4.4\times 10^{15}$~(erg~s$^{-1}$~g$^{-1}$) and is partly radiated away and balanced by the pressure work within a very narrow region, where the non-thermal electron energy is mostly deposited (see bottom row of Figure~\ref{evol_atm}). The electron heating in the lower atmosphere is negligible in comparison with the other terms and it is mostly radiated away.

As the atmosphere evolves, most of the conductive energy in the transition region is radiated away and very little is left to heat the plasma. This energy exchange occurs in a very narrow layer, less than 10~m thick, at $\approx$~0.866~Mm, where the conductive heating rate increases from almost 0 to $5~\times~10^{17}$~erg~s$^{-1}$~g$^{-1}$ (see the inset in Figure~\ref{energy_terms_balance_run2}(c)). The atmosphere is already ionized in this region and does not radiate efficiently anymore, thus increasing the total internal energy and temperature. Note that for a better comparison, the last two columns of Figure~\ref{energy_terms_balance_run2} do not show the total length of the peak. 

As shown in the bottom panels, most of the total energy gain (black thick line) is used to increase the thermal energy (red line), as in the case of PL. However, these energy change rates are one to two orders of magnitude higher than those in PL.

    \section{B. Effects of the amplitude of the electron flux $\mathfrak{F}_{max}$}\label{sect:results_flux}
Here we present the result of SA runs of different peak energy fluxes of the non-thermal electrons, $\mathfrak{F}_{max}$, but with the same triangular-shaped temporal profile of $\mathfrak{F}(t)$ as in Runs SA1 and PL. The characteristics of these runs are shown in Table~\ref{table_runs}. As expected, a higher electron flux leads to higher maximum upflow and downflow speeds and temperatures which occur at earlier times. The only exception is the maximum temperature that almost always occurs at the time of $\mathfrak{F}_{max}$, i.e., $t=60$~s. A factor of 20 increase in $\mathfrak{F}_{max}$ from $5.7 \times 10^8$~erg~s$^{-1}$~cm$^{-2}$ (Run~SA3) to $1.2 \times 10^{10}$~erg~s$^{-1}$~cm$^{-2}$ (Run~SA1) only results in a factor of two increase in the maximum upflow speed $v_{\rm max}$ from 377 to 750~km~s$^{-1}$, a factor of 50\% increase in the maximum downflow speed $v_{\rm min}$ from $-28$ to $-41$~km~s$^{-1}$, and a factor of three increase in the maximum temperature from 8 to 23~MK. These results are in line with those of \citealt[][see his Table~7.1]{2008sfpa.book.....L} for simulations using the same spectrum of stochastically accelerated electrons as we have adopted here.

\begin{figure*}[tbh]
\centering
  \epsscale{1.15}
  \plotone{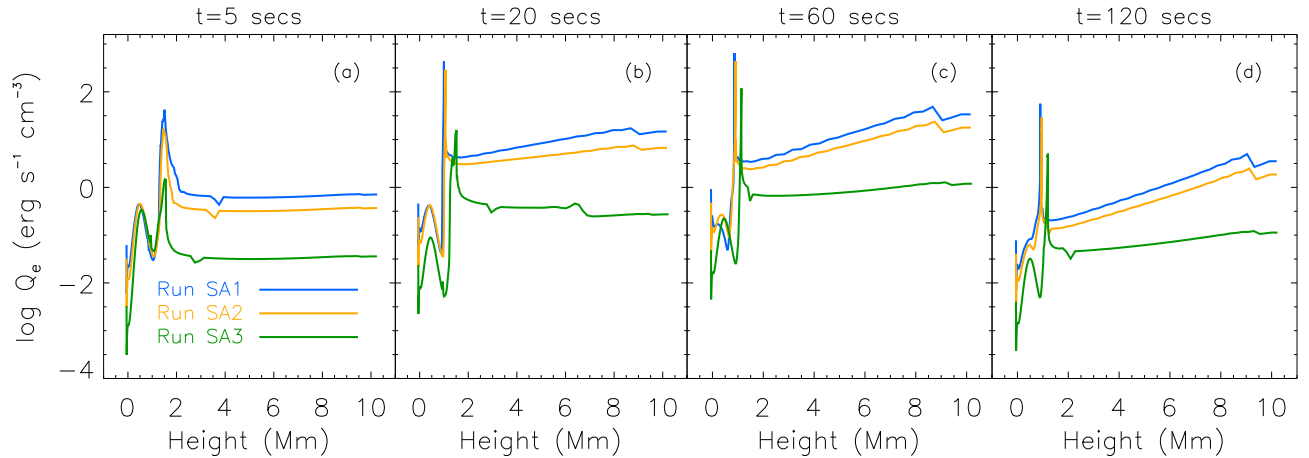}
  \caption{Spatial distribution of the electron heating rate $Q_e(z)$ for Runs~SA1, SA2 and SA3 (see Table~\ref{table_runs}) at selected times, $t=5$, 20, 60 and 120~s. [A color version of this figure is available in the online journal.].}
     \label{bflux_runs1_2_3}
\end{figure*}

\citet{1985ApJ...289..414F} found that an electron energy flux of $< 10^{10}$~erg~s$^{-1}$~cm$^{-2}$ can produce so-called ``gentle" chromospheric evaporation with upflow speeds of tens of km~s$^{-1}$, while an energy flux of $> 3 \times 10^{10}$~erg~s$^{-1}$~cm$^{-2}$ would produce ``explosive" evaporation with upflow speeds of hundreds of km~s$^{-1}$ and downflow speeds of tens of km~s$^{-1}$. As shown in Table~\ref{table_runs}, all the SA runs have energy flux below this threshold and yet all exhibit characteristics of ``explosive" evaporation. This demonstrates that the energy flux is not the only factor that dictates the atmospheric response to electron heating; the electron spectral shape is also critical because it determines the spatial distribution of the energy deposition. The key difference is that \citet{1985ApJ...289..414F} injected a single power-law electron spectrum of an index of $\delta=4$ and cutoff energy of 20~keV, which is comparable to that in our PL run. Such non-thermal electrons directly deposit most of their energy deep in the chromosphere, which is largely radiated away (see Figure~\ref{energy_terms_balance_run1}). In our SA runs, the electron spectrum has a prominent quasi-thermal component and thus results in heating primarily in the corona (in terms of heating rate per unit mass, see Figure~\ref{energy_terms_balance_run2}). Then thermal conduction becomes the primary heating agent in the lower atmosphere, mainly in the transition region where radiative loss is not as strong as in the chromosphere. As a result, there is relatively more energy left to heat the plasma and drive chromospheric evaporation. This is similar to the case in Paper~I, although radiative transfer was not included there. These result indicates that the stochastically accelerated electrons, because of their preferential heating in the corona rather than directly in the chromosphere, are more efficient in driving chromospheric evaporation than power-law electrons with a cutoff energy. This also indicates that, in addition to direct collisional heating by non-thermal electrons as commonly believed, thermal conduction can play an important role in driving chromospheric evaporation, as observed in some flares \citep[e.g.,][]{1988ApJ...329..456Z, 2009A&A...498..891B}. This is even more so for weak flares, such our SA3 run.

In the case of faint flares (Run~SA3) most of the electron heating rate, $Q_e$,  is transformed into conductive heating and radiated away along the atmosphere, explaining why the electron heating in Figure~\ref{bflux_runs1_2_3} (green line) remains almost constant in the corona. By increasing the electron flux $\mathfrak{F}$, in Runs~SA1 and SA2 (blue and yellow lines), the heating increases at the top of the loop with time.

    \subsection{B1.~Lines emission}\label{sect:emission_lines_flx}
\begin{figure}[tbh]
\centering
  \epsscale{0.6}
  \plotone{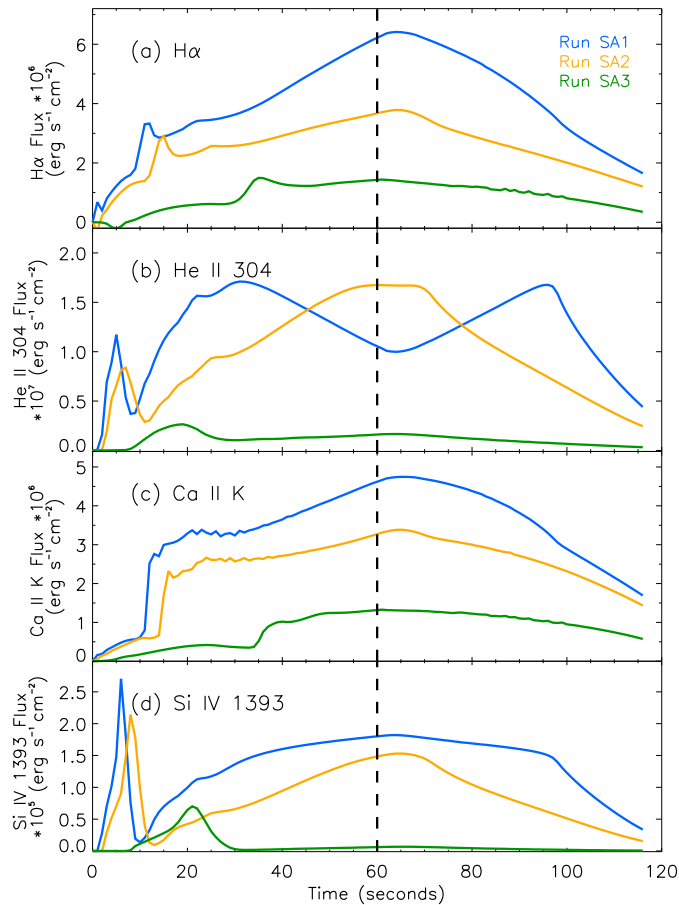}
  \caption{Light curves of the H$\alpha$~6563~\AA, \ion{He}{2}~304~\AA, \ion{Ca}{2}~K~3934~\AA, and \ion{Si}{4}~1393~\AA\ lines for Runs~SA1--SA3. The emission at the initial time has been subtracted. The vertical dashed line indicates $t_{max}$. [A color version of this figure is available in the online journal.].}
     \label{light_curve_flx}
\end{figure} 

Figure~\ref{light_curve_flx} shows the light curve of the H$\alpha$, \ion{He}{2}~304~\AA, \ion{Ca}{2}~K and \ion{Si}{4}~1393~\AA\ lines for the three SA models, where the emission of the quiet Sun (at the initial time) has been removed.
In general, for higher peak electron energy fluxes, the line intensity increases faster with time, because of more rapid hydrodynamic response of the atmosphere, and reaches higher peak values. This is the case for the H$\alpha$~and \ion{Ca}{2}~K fluxes, which are positively correlated with the electron flux, peaking around $t_{max}$. However, the \ion{He}{2}~304~\AA\ intensity in Run~SA1 decreases in the middle portion of the flare duration, showing anti-correlation in time with the electron flux. Each of the \ion{He}{2} light curves of the three runs also shows distinct behaviors.
In Table~\ref{table_runs} we can see that the maximum temperature decreases linearly with the flux of the injected electrons. Therefore the recombination-photoionization processes may play a less important role in the formation of the line, giving preference to the collision excitation processes. This may explain why the shape of the \ion{He}{2}~304~\AA\ line profiles and the flux changes with $\mathfrak{F}$. 

The temporal evolution of the \ion{Si}{4}~1393~\AA\ flux shows an earlier bump associated with the increase of the electron density in the chromosphere and transition region. As discussed in Section~\ref{sect:Si_line_emission}, an increase of the electron density leads to an increase of the emission measure at this wavelength range and thus the increase in intensity of this optically-thin line.

\bibliographystyle{apj}
\bibliography{ads}

\begin{thebibliography}{72}
\expandafter\ifx\csname natexlab\endcsname\relax\def\natexlab#1{#1}\fi

\bibitem[{{Abbett} \& {Hawley}(1999)}]{1999ApJ...521..906A}
{Abbett}, W.~P. \& {Hawley}, S.~L. 1999, \apj, 521, 906

\bibitem[{{Allred} {et~al.}(2015){Allred}, {Kowalski}, \&
  {Carlsson}}]{2015TESS....130207A}
{Allred}, J., {Kowalski}, A., \& {Carlsson}, M. 2015, in AAS/AGU Triennial
  Earth-Sun Summit, Vol.~1, AAS/AGU Triennial Earth-Sun Summit, 30207

\bibitem[{{Allred} {et~al.}(2005){Allred}, {Hawley}, {Abbett}, \&
  {Carlsson}}]{2005ApJ...630..573A}
{Allred}, J.~C., {Hawley}, S.~L., {Abbett}, W.~P., \& {Carlsson}, M. 2005,
  \apj, 630, 573

\bibitem[{{Andretta} {et~al.}(2003){Andretta}, {Del Zanna}, \&
  {Jordan}}]{2003A&A...400..737A}
{Andretta}, V., {Del Zanna}, G., \& {Jordan}, S.~D. 2003, \aap, 400, 737

\bibitem[{{Battaglia} {et~al.}(2009){Battaglia}, {Fletcher}, \&
  {Benz}}]{2009A&A...498..891B}
{Battaglia}, M., {Fletcher}, L., \& {Benz}, A.~O. 2009, \aap, 498, 891

\bibitem[{{Carlsson} \& {Stein}(1992)}]{1992ApJ...397L..59C}
{Carlsson}, M. \& {Stein}, R.~F. 1992, \apjl, 397, L59

\bibitem[{{Carlsson} \& {Stein}(1997)}]{1997ApJ...481..500C}
---. 1997, \apj, 481, 500

\bibitem[{{Cauzzi} {et~al.}(1993){Cauzzi}, {Falchi}, {Falciani}, \&
  {Smaldone}}]{1993AdSpR..13..311C}
{Cauzzi}, G., {Falchi}, A., {Falciani}, R., \& {Smaldone}, L.~A. 1993, Advances
  in Space Research, 13, 311

\bibitem[{{Chen} \& {Petrosian}(2013)}]{2013ApJ...777...33C}
{Chen}, Q. \& {Petrosian}, V. 2013, \apj, 777, 33

\bibitem[{{De Pontieu} {et~al.}(2015){De Pontieu}, {McIntosh},
  {Martinez-Sykora}, {Peter}, \& {Pereira}}]{2015ApJ...799L..12D}
{De Pontieu}, B., {McIntosh}, S., {Martinez-Sykora}, J., {Peter}, H., \&
  {Pereira}, T.~M.~D. 2015, \apjl, 799, L12

\bibitem[{{De Pontieu} {et~al.}(2014){De Pontieu}, {Title}, {Lemen}, {Kushner},
  {Akin}, {Allard}, {Berger}, {Boerner}, {Cheung}, {Chou}, {Drake}, {Duncan},
  {Freeland}, {Heyman}, {Hoffman}, {Hurlburt}, {Lindgren}, {Mathur}, {Rehse},
  {Sabolish}, {Seguin}, {Schrijver}, {Tarbell}, {W{\"u}lser}, {Wolfson},
  {Yanari}, {Mudge}, {Nguyen-Phuc}, {Timmons}, {van Bezooijen}, {Weingrod},
  {Brookner}, {Butcher}, {Dougherty}, {Eder}, {Knagenhjelm}, {Larsen},
  {Mansir}, {Phan}, {Boyle}, {Cheimets}, {DeLuca}, {Golub}, {Gates}, {Hertz},
  {McKillop}, {Park}, {Perry}, {Podgorski}, {Reeves}, {Saar}, {Testa}, {Tian},
  {Weber}, {Dunn}, {Eccles}, {Jaeggli}, {Kankelborg}, {Mashburn}, {Pust},
  {Springer}, {Carvalho}, {Kleint}, {Marmie}, {Mazmanian}, {Pereira}, {Sawyer},
  {Strong}, {Worden}, {Carlsson}, {Hansteen}, {Leenaarts}, {Wiesmann},
  {Aloise}, {Chu}, {Bush}, {Scherrer}, {Brekke}, {Martinez-Sykora}, {Lites},
  {McIntosh}, {Uitenbroek}, {Okamoto}, {Gummin}, {Auker}, {Jerram}, {Pool}, \&
  {Waltham}}]{2014SoPh..289.2733D}
{De Pontieu}, B., {Title}, A.~M., {Lemen}, J.~R., {Kushner}, G.~D., {Akin},
  D.~J., {Allard}, B., {Berger}, T., {Boerner}, P., {Cheung}, M., {Chou}, C.,
  {Drake}, J.~F., {Duncan}, D.~W., {Freeland}, S., {Heyman}, G.~F., {Hoffman},
  C., {Hurlburt}, N.~E., {Lindgren}, R.~W., {Mathur}, D., {Rehse}, R.,
  {Sabolish}, D., {Seguin}, R., {Schrijver}, C.~J., {Tarbell}, T.~D.,
  {W{\"u}lser}, J.-P., {Wolfson}, C.~J., {Yanari}, C., {Mudge}, J.,
  {Nguyen-Phuc}, N., {Timmons}, R., {van Bezooijen}, R., {Weingrod}, I.,
  {Brookner}, R., {Butcher}, G., {Dougherty}, B., {Eder}, J., {Knagenhjelm},
  V., {Larsen}, S., {Mansir}, D., {Phan}, L., {Boyle}, P., {Cheimets}, P.~N.,
  {DeLuca}, E.~E., {Golub}, L., {Gates}, R., {Hertz}, E., {McKillop}, S.,
  {Park}, S., {Perry}, T., {Podgorski}, W.~A., {Reeves}, K., {Saar}, S.,
  {Testa}, P., {Tian}, H., {Weber}, M., {Dunn}, C., {Eccles}, S., {Jaeggli},
  S.~A., {Kankelborg}, C.~C., {Mashburn}, K., {Pust}, N., {Springer}, L.,
  {Carvalho}, R., {Kleint}, L., {Marmie}, J., {Mazmanian}, E., {Pereira},
  T.~M.~D., {Sawyer}, S., {Strong}, J., {Worden}, S.~P., {Carlsson}, M.,
  {Hansteen}, V.~H., {Leenaarts}, J., {Wiesmann}, M., {Aloise}, J., {Chu},
  K.-C., {Bush}, R.~I., {Scherrer}, P.~H., {Brekke}, P., {Martinez-Sykora}, J.,
  {Lites}, B.~W., {McIntosh}, S.~W., {Uitenbroek}, H., {Okamoto}, T.~J.,
  {Gummin}, M.~A., {Auker}, G., {Jerram}, P., {Pool}, P., \& {Waltham}, N.
  2014, \solphys, 289, 2733

\bibitem[{{Dere} {et~al.}(1997){Dere}, {Landi}, {Mason}, {Monsignori Fossi}, \&
  {Young}}]{1997A&AS..125..149D}
{Dere}, K.~P., {Landi}, E., {Mason}, H.~E., {Monsignori Fossi}, B.~C., \&
  {Young}, P.~R. 1997, \aaps, 125, 149

\bibitem[{{Donea}(2011)}]{2011SSRv..158..451D}
{Donea}, A. 2011, \ssr, 158, 451

\bibitem[{{Dorfi} \& {Drury}(1987)}]{1987JCoPh..69..175D}
{Dorfi}, E.~A. \& {Drury}, L.~O. 1987, Journal of Computational Physics, 69,
  175

\bibitem[{{Emslie}(1978)}]{1978ApJ...224..241E}
{Emslie}, A.~G. 1978, \apj, 224, 241

\bibitem[{{Emslie}(1981)}]{1981ApJ...249..817E}
---. 1981, \apj, 249, 817

\bibitem[{{Fisher} {et~al.}(2012){Fisher}, {Bercik}, {Welsch}, \&
  {Hudson}}]{2012SoPh..277...59F}
{Fisher}, G.~H., {Bercik}, D.~J., {Welsch}, B.~T., \& {Hudson}, H.~S. 2012,
  \solphys, 277, 59

\bibitem[{{Fisher} {et~al.}(1985){Fisher}, {Canfield}, \&
  {McClymont}}]{1985ApJ...289..414F}
{Fisher}, G.~H., {Canfield}, R.~C., \& {McClymont}, A.~N. 1985, \apj, 289, 414

\bibitem[{{Golding} {et~al.}(2014){Golding}, {Carlsson}, \&
  {Leenaarts}}]{2014ApJ...784...30G}
{Golding}, T.~P., {Carlsson}, M., \& {Leenaarts}, J. 2014, \apj, 784, 30

\bibitem[{{Gustafsson}(1973)}]{Gustafsson}
{Gustafsson}, B. 1973, A FORTRAN Program for Calculating 'Continuous'
  Absorption Coefficients of Stellar Atmospheres (Uppsala: Lalidstingets
  Verkstader)

\bibitem[{{Hamilton} \& {Petrosian}(1992)}]{1992ApJ...398..350H}
{Hamilton}, R.~J. \& {Petrosian}, V. 1992, \apj, 398, 350

\bibitem[{{Holman}(1985)}]{1985ApJ...293..584H}
{Holman}, G.~D. 1985, \apj, 293, 584

\bibitem[{{Hoyng} {et~al.}(1981){Hoyng}, {Duijveman}, {Machado}, {Rust},
  {Svestka}, {Boelee}, {de Jager}, {Frost}, {Lafleur}, {Simnett}, {van Beek},
  \& {Woodgate}}]{1981ApJ...246L.155H}
{Hoyng}, P., {Duijveman}, A., {Machado}, M.~E., {Rust}, D.~M., {Svestka}, Z.,
  {Boelee}, A., {de Jager}, C., {Frost}, K.~T., {Lafleur}, H., {Simnett},
  G.~M., {van Beek}, H.~F., \& {Woodgate}, B.~E. 1981, \apjl, 246, L155

\bibitem[{{Johannesson} {et~al.}(1995){Johannesson}, {Marquette}, \&
  {Zirin}}]{1995SoPh..161..201J}
{Johannesson}, A., {Marquette}, W., \& {Zirin}, H. 1995, \solphys, 161, 201

\bibitem[{{Johannesson} {et~al.}(1998){Johannesson}, {Marquette}, \&
  {Zirin}}]{1998SoPh..177..265J}
{Johannesson}, A., {Marquette}, W.~H., \& {Zirin}, H. 1998, \solphys, 177, 265

\bibitem[{{Jordan}(1975)}]{1975MNRAS.170..429J}
{Jordan}, C. 1975, \mnras, 170, 429

\bibitem[{{Jordan} {et~al.}(1993){Jordan}, {Thompson}, {Thomas}, \&
  {Neupert}}]{1993ApJ...406..346J}
{Jordan}, S.~D., {Thompson}, W.~T., {Thomas}, R.~J., \& {Neupert}, W.~M. 1993,
  \apj, 406, 346

\bibitem[{{Ka{\v s}parov{\'a}} {et~al.}(2009){Ka{\v s}parov{\'a}}, {Varady},
  {Heinzel}, {Karlick{\'y}}, \& {Moravec}}]{2009A&A...499..923K}
{Ka{\v s}parov{\'a}}, J., {Varady}, M., {Heinzel}, P., {Karlick{\'y}}, M., \&
  {Moravec}, Z. 2009, \aap, 499, 923

\bibitem[{{Kosovichev}(2007)}]{2007ApJ...670L..65K}
{Kosovichev}, A.~G. 2007, \apjl, 670, L65

\bibitem[{{Kosovichev} \& {Zharkova}(1998)}]{1998Natur.393..317K}
{Kosovichev}, A.~G. \& {Zharkova}, V.~V. 1998, \nat, 393, 317

\bibitem[{{Kundu} {et~al.}(1994){Kundu}, {White}, {Gopalswamy}, \&
  {Lim}}]{1994ApJS...90..599K}
{Kundu}, M.~R., {White}, S.~M., {Gopalswamy}, N., \& {Lim}, J. 1994, \apjs, 90,
  599

\bibitem[{{Laming} \& {Feldman}(1992)}]{1992ApJ...386..364L}
{Laming}, J.~M. \& {Feldman}, U. 1992, \apj, 386, 364

\bibitem[{{Landi} {et~al.}(2013){Landi}, {Young}, {Dere}, {Del Zanna}, \&
  {Mason}}]{2013ApJ...763...86L}
{Landi}, E., {Young}, P.~R., {Dere}, K.~P., {Del Zanna}, G., \& {Mason}, H.~E.
  2013, \apj, 763, 86

\bibitem[{{Leach} \& {Petrosian}(1981)}]{1981ApJ...251..781L}
{Leach}, J. \& {Petrosian}, V. 1981, \apj, 251, 781

\bibitem[{{Leenaarts} {et~al.}(2012){Leenaarts}, {Carlsson}, \& {Rouppe van der
  Voort}}]{2012ApJ...749..136L}
{Leenaarts}, J., {Carlsson}, M., \& {Rouppe van der Voort}, L. 2012, \apj, 749,
  136

\bibitem[{{Leenaarts} {et~al.}(2006){Leenaarts}, {Rutten}, {S{\"u}tterlin},
  {Carlsson}, \& {Uitenbroek}}]{2006A&A...449.1209L}
{Leenaarts}, J., {Rutten}, R.~J., {S{\"u}tterlin}, P., {Carlsson}, M., \&
  {Uitenbroek}, H. 2006, \aap, 449, 1209

\bibitem[{{Lin}(1985)}]{1985SoPh..100..537L}
{Lin}, R.~P. 1985, \solphys, 100, 537

\bibitem[{{Lindsey} \& {Donea}(2008)}]{2008SoPh..251..627L}
{Lindsey}, C. \& {Donea}, A.-C. 2008, \solphys, 251, 627

\bibitem[{{Liu} {et~al.}(2004){Liu}, {Petrosian}, \&
  {Mason}}]{2004ApJ...613L..81L}
{Liu}, S., {Petrosian}, V., \& {Mason}, G.~M. 2004, \apjl, 613, L81

\bibitem[{{Liu} {et~al.}(2006){Liu}, {Petrosian}, \&
  {Mason}}]{2006ApJ...636..462L}
---. 2006, \apj, 636, 462

\bibitem[{{Liu}(2008)}]{2008sfpa.book.....L}
{Liu}, W. 2008, {Solar Flares as Natural Particle Accelerators: A High-energy
  View from X-ray Observations and Theoretical Models} (VDM Verlag Dr)

\bibitem[{{Liu} {et~al.}(2013){Liu}, {Chen}, \&
  {Petrosian}}]{2013ApJ...767..168L}
{Liu}, W., {Chen}, Q., \& {Petrosian}, V. 2013, \apj, 767, 168

\bibitem[{{Liu} {et~al.}(2009){Liu}, {Petrosian}, \&
  {Mariska}}]{2009ApJ...702.1553L}
{Liu}, W., {Petrosian}, V., \& {Mariska}, J.~T. 2009, \apj, 702, 1553

\bibitem[{{Mariska} {et~al.}(1989){Mariska}, {Emslie}, \&
  {Li}}]{1989ApJ...341.1067M}
{Mariska}, J.~T., {Emslie}, A.~G., \& {Li}, P. 1989, \apj, 341, 1067

\bibitem[{{McTiernan} \& {Petrosian}(1990)}]{1990ApJ...359..524M}
{McTiernan}, J.~M. \& {Petrosian}, V. 1990, \apj, 359, 524

\bibitem[{{Milkey} \& {Mihalas}(1973)}]{1973ApJ...185..709M}
{Milkey}, R.~W. \& {Mihalas}, D. 1973, \apj, 185, 709

\bibitem[{{Miller}(1997)}]{1997ApJ...491..939M}
{Miller}, J.~A. 1997, \apj, 491, 939

\bibitem[{{Milligan} \& {Dennis}(2009)}]{2009ApJ...699..968M}
{Milligan}, R.~O. \& {Dennis}, B.~R. 2009, \apj, 699, 968

\bibitem[{{O'Dwyer} {et~al.}(2010){O'Dwyer}, {Del Zanna}, {Mason}, {Weber}, \&
  {Tripathi}}]{2010A&A...521A..21O}
{O'Dwyer}, B., {Del Zanna}, G., {Mason}, H.~E., {Weber}, M.~A., \& {Tripathi},
  D. 2010, \aap, 521, A21

\bibitem[{{Olluri} {et~al.}(2013){Olluri}, {Gudiksen}, \&
  {Hansteen}}]{2013ApJ...767...43O}
{Olluri}, K., {Gudiksen}, B.~V., \& {Hansteen}, V.~H. 2013, \apj, 767, 43

\bibitem[{{Paletou} {et~al.}(1993){Paletou}, {Vial}, \&
  {Auer}}]{1993A&A...274..571P}
{Paletou}, F., {Vial}, J.-C., \& {Auer}, L.~H. 1993, \aap, 274, 571

\bibitem[{{Park} \& {Petrosian}(1995)}]{1995ApJ...446..699P}
{Park}, B.~T. \& {Petrosian}, V. 1995, \apj, 446, 699

\bibitem[{{Petrosian}(2012)}]{2012SSRv..173..535P}
{Petrosian}, V. 2012, \ssr, 173, 535

\bibitem[{{Petrosian} \& {Chen}(2010)}]{2010ApJ...712L.131P}
{Petrosian}, V. \& {Chen}, Q. 2010, \apjl, 712, L131

\bibitem[{{Petrosian} \& {Donaghy}(1999)}]{1999ApJ...527..945P}
{Petrosian}, V. \& {Donaghy}, T.~Q. 1999, \apj, 527, 945

\bibitem[{{Petrosian} {et~al.}(2002){Petrosian}, {Donaghy}, \&
  {McTiernan}}]{2002ApJ...569..459P}
{Petrosian}, V., {Donaghy}, T.~Q., \& {McTiernan}, J.~M. 2002, \apj, 569, 459

\bibitem[{{Petrosian} \& {East}(2008)}]{2008ApJ...682..175P}
{Petrosian}, V. \& {East}, W.~E. 2008, \apj, 682, 175

\bibitem[{{Petrosian} \& {Liu}(2004)}]{2004ApJ...610..550P}
{Petrosian}, V. \& {Liu}, S. 2004, \apj, 610, 550

\bibitem[{{P{\"o}tzi} {et~al.}(2013){P{\"o}tzi}, {Temmer}, {Veronig},
  {Hirtenfellner-Polanec}, \& {Baumgartner}}]{2013EGUGA..15.1459P}
{P{\"o}tzi}, W., {Temmer}, M., {Veronig}, A., {Hirtenfellner-Polanec}, W., \&
  {Baumgartner}, D. 2013, in EGU General Assembly Conference Abstracts,
  Vol.~15, EGU General Assembly Conference Abstracts, 1459

\bibitem[{{Ramaty}(1979)}]{1979AIPC...56..135R}
{Ramaty}, R. 1979, in American Institute of Physics Conference Series, Vol.~56,
  Particle Acceleration Mechanisms in Astrophysics, ed. J.~{Arons}, C.~{McKee},
  \& C.~{Max}, 135--154

\bibitem[{{Reep} {et~al.}(2015){Reep}, {Bradshaw}, \&
  {Alexander}}]{2015arXiv150608115R}
{Reep}, J., {Bradshaw}, S., \& {Alexander}, D. 2015, ArXiv e-prints

\bibitem[{{Rubio da Costa} {et~al.}(2015){Rubio da Costa}, {Kleint},
  {Petrosian}, {Sainz Dalda}, \& {Liu}}]{2015ApJ...804...56R}
{Rubio da Costa}, F., {Kleint}, L., {Petrosian}, V., {Sainz Dalda}, A., \&
  {Liu}, W. 2015, \apj, 804, 56

\bibitem[{{Sakao}(1994)}]{1994PhDT.......335S}
{Sakao}, T. 1994, PhD thesis, (University of Tokyo), (1994)

\bibitem[{{Sironi} \& {Spitkovsky}(2009)}]{2009ApJ...707L..92S}
{Sironi}, L. \& {Spitkovsky}, A. 2009, \apjl, 707, L92

\bibitem[{{Testa} {et~al.}(2014){Testa}, {De Pontieu}, {Allred}, {Carlsson},
  {Reale}, {Daw}, {Hansteen}, {Martinez-Sykora}, {Liu}, {DeLuca}, {Golub},
  {McKillop}, {Reeves}, {Saar}, {Tian}, {Lemen}, {Title}, {Boerner},
  {Hurlburt}, {Tarbell}, {Wuelser}, {Kleint}, {Kankelborg}, \&
  {Jaeggli}}]{2014Sci...346B.315T}
{Testa}, P., {De Pontieu}, B., {Allred}, J., {Carlsson}, M., {Reale}, F.,
  {Daw}, A., {Hansteen}, V., {Martinez-Sykora}, J., {Liu}, W., {DeLuca}, E.~E.,
  {Golub}, L., {McKillop}, S., {Reeves}, K., {Saar}, S., {Tian}, H., {Lemen},
  J., {Title}, A., {Boerner}, P., {Hurlburt}, N., {Tarbell}, T.~D., {Wuelser},
  J.~P., {Kleint}, L., {Kankelborg}, C., \& {Jaeggli}, S. 2014, Science, 346,
  B315

\bibitem[{{Tsuneta} \& {Naito}(1998)}]{1998ApJ...495L..67T}
{Tsuneta}, S. \& {Naito}, T. 1998, \apjl, 495, L67

\bibitem[{{{\v S}vestka}(1976)}]{1976saop.book..141S}
{{\v S}vestka}, Z. 1976, {Several Solar Aspects of Flare-Associated Particle
  Events}, ed. P.~S. {McIntosh} \& M.~{Dryer}, 141--162

\bibitem[{{Vial}(1982)}]{1982ApJ...253..330V}
{Vial}, J.~C. 1982, \apj, 253, 330

\bibitem[{{Woods} {et~al.}(2012){Woods}, {Eparvier}, {Hock}, {Jones},
  {Woodraska}, {Judge}, {Didkovsky}, {Lean}, {Mariska}, {Warren}, {McMullin},
  {Chamberlin}, {Berthiaume}, {Bailey}, {Fuller-Rowell}, {Sojka}, {Tobiska}, \&
  {Viereck}}]{2012SoPh..275..115W}
{Woods}, T.~N., {Eparvier}, F.~G., {Hock}, R., {Jones}, A.~R., {Woodraska}, D.,
  {Judge}, D., {Didkovsky}, L., {Lean}, J., {Mariska}, J., {Warren}, H.,
  {McMullin}, D., {Chamberlin}, P., {Berthiaume}, G., {Bailey}, S.,
  {Fuller-Rowell}, T., {Sojka}, J., {Tobiska}, W.~K., \& {Viereck}, R. 2012,
  \solphys, 275, 115

\bibitem[{{Zarro} \& {Lemen}(1988)}]{1988ApJ...329..456Z}
{Zarro}, D.~M. \& {Lemen}, J.~R. 1988, \apj, 329, 456

\bibitem[{{Zharkova}(2008)}]{2008SoPh..251..641Z}
{Zharkova}, V.~V. 2008, \solphys, 251, 641

\bibitem[{{Zirin}(1988)}]{1988assu.book.....Z}
{Zirin}, H. 1988, {Astrophysics of the sun}

\end{thebibliography}
\end{document}